\documentclass[twocolumn]{aastex61hack}

\newcommand{\rii}{\emph{r}-II}
\newcommand{\rp}{\emph{r}-process}
\newcommand{\kt}{K\&T}

\usepackage{enumerate}
\usepackage{graphicx}
\usepackage{amsmath}
\usepackage{natbib}

\begin{document}

\title{Actinide Production in the Neutron-Rich Ejecta \\
of a Neutron Star Merger}

\author{Erika M.\ Holmbeck}
\affiliation{Department of Physics, University of Notre Dame, Notre Dame, IN 46556, USA}
\affiliation{JINA Center for the Evolution of the Elements, USA}

\author{Trevor M.\ Sprouse}
\affiliation{Department of Physics, University of Notre Dame, Notre Dame, IN 46556, USA}

\author{Matthew R.\ Mumpower}
\affiliation{Theoretical Division, Los Alamos National Laboratory, Los Alamos, NM 87545, USA}
\affiliation{JINA Center for the Evolution of the Elements, USA}

\author{Nicole Vassh}
\affiliation{Department of Physics, University of Notre Dame, Notre Dame, IN 46556, USA}

\author{Rebecca Surman}
\affiliation{Department of Physics, University of Notre Dame, Notre Dame, IN 46556, USA}
\affiliation{JINA Center for the Evolution of the Elements, USA}

\author{Timothy C.\ Beers}
\affiliation{Department of Physics, University of Notre Dame, Notre Dame, IN 46556, USA}
\affiliation{JINA Center for the Evolution of the Elements, USA}

\author{Toshihiko Kawano}
\affiliation{Theoretical Division, Los Alamos National Laboratory, Los Alamos, NM 87545, USA}

\correspondingauthor{Erika M.\ Holmbeck}
\email{eholmbec@nd.edu}

\begin{abstract}
%

The rapid-neutron-capture (``\emph{r}-") process is responsible for synthesizing many of the heavy elements observed in both the Solar System and Galactic metal-poor halo stars.
Simulations of \rp\ nucleosynthesis can reproduce abundances derived from observations with varying success, but so far fail to account for the observed over-enhancement of actinides, present in about 30\% of \rp-enhanced stars.
In this work, we investigate actinide production in the dynamical ejecta of a neutron star merger and explore if varying levels of neutron richness can reproduce the actinide boost.
We also investigate the sensitivity of actinide production on nuclear physics properties: fission distribution, $\beta$-decay, and mass model.
For most cases, the actinides are over-produced in our models if the initial conditions are sufficiently neutron-rich for fission cycling.
We find that actinide production can be so robust in the dynamical ejecta that an additional lanthanide-rich, actinide-poor component is necessary in order to match observations of actinide-boost stars. We present a simple actinide-dilution model that folds in estimated contributions from two nucleosynthetic sites within a merger event. Our study suggests that while the dynamical ejecta of a neutron star merger are likely production sites for the formation of actinides, a significant contribution from another site or sites (e.g., the neutron star merger accretion disk wind) is required to explain abundances of \rp-enhanced, metal-poor stars.
\end{abstract}


\received{}
\revised{}
\submitjournal{\apj}

\section{Introduction}

Of the elements heavier than iron observable in the Solar System, the rapid neutron-capture (``\emph{r}-") process accounts for the formation of roughly half.
Although the physical mechanism responsible for \rp\ nucleosynthesis has been known since \citet{b2fh} and \citet{cameron1957}, its astrophysical site has been a long-standing question \citep{Arnould+07,Thielemann+11}.

One way to observationally investigate the \rp\ site is to study metal-poor stars with nearly pristine atmospheres. Of all stars below a metallicity of $[{\rm Fe/H}]\footnote{$[{\rm A/B}] = \log(N_A/N_B)_* - \log(N_A/N_B)_\sun$, where $N$ is the number density of an element in the star (*) compared to the Sun ($\sun$).}=-2$, about 15\% show evidence in their photospheres of \rp-enhancement \citep{barklem2005}. These are characterized according to their europium-to-iron ratio as ``\emph{r}-I" (+$0.3 \leq {\rm [Eu/Fe]} \leq +1.0$) and ``\rii" \citep[${\rm [Eu/Fe]} > +1.0$;][]{beers2005} stars. The \emph{r}-I and \rii\ stars are laboratories for studying nearly-pure \rp\ events, since the photospheres of giant stars retain---with the exception of some of the lighter elements---the elemental abundance ratio of the prenatal cloud from which they were created.

If metallicity is taken as a reliable proxy for age, e.g., \citet{Piatti+17}, Milky Way \emph{r}-I and \rii\ stars are estimated to be about $\sim$11--12 Gyr old, requiring \rp\ production to operate early in Galactic history \citep{sneden2008,Roederer+14}. The astrophysical site seemingly most compatible with early enrichment is core-collapse supernovae (SNe), considered natural \rp\ laboratories since \citet{b2fh} \citep{Meyer+92,Woosley+94}. However, two major recent observations challenge this assumption. 

The discovery of seven \rii\ stars in the dwarf spheroidal galaxy Reticulum II \citep[``Ret II";][]{ji2016,roederer2016} indicates copious enrichment by a single event, rather than the slow buildup that would be expected from SNe. A likely candidate for this event is a neutron star merger (NSM). The prompt, very neutron-rich ejecta from NSM events have long been considered attractive environments for the \rp\ \citep{lattimer1974,Meyer89,Freiburghaus+99,Goriely+11}. However, the assumed long coalescence timescales of hundreds to thousands of Myr were thought to disfavor NSM as sources of observed low metallicity \rp\ enrichments \citep{Mathews+90,Argast+04}. The Ret II observations offer a path to relieve this tension, as updated Galactic chemical evolution simulations show long coalescence timescales can be accommodated if the Milky Way halo is formed entirely or in part by the accretion of low-metallicity dwarf spheroidal galaxies \citep{Hirai+15}. 

The suggestion that NSMs are viable \rp\ sites received dramatic confirmation in 2017 with the observation of gravitational wave event GW170817 \citep{abbott2017}. The GW170817 NSM produced radiation across the electromagnetic spectrum \citep{drout2017,shappee2017}, consistent with the ejection of some hundredths of a Solar mass of lanthanide-rich material \citep{kilpatrick2017,cowperthwaite2017}. Though the uncertainties are still large, the estimated amount of \rp\ material produced by this event appears consistent both with that required for the Ret II enrichments and to explain the bulk of the heavy \rp\ abundances in the Galaxy \citep{Cote+18}. 

Among the \rp-enhanced stars, an even rarer signature is found: the ``actinide boost," present in about 30\% of \rp-enhanced stars \citep{mashonkina2014}.
The actinides $^{232}$Th and $^{238}$U are produced exclusively via the \rp, and their long half-lives, 14.0 Gyr and 4.468 Gyr, respectively, allow for their potential use as cosmochronometers \citep{Cowan+91}.
$^{235}$U is also made by the \rp, but with a shorter half-life of 0.704 Gyr, it does not contribute significantly to the total uranium abundance in metal-poor stars.
The ages of \rp\ events can be extracted from the observed abundance ratios of actinides to co-produced stable species, given some estimate of their initial production. 
Although the comparison can be to any stable \rp\ element, it is common to use Eu because of its role in quantifying \rp\ enrichment of metal-poor stars.
Observed thorium-to-europium abundance ratios in most cases result in inferred ages of $\sim$2--14 Gyr; see Figure~9 of \citet{mashonkina2014}. In contrast, actinide-boost stars are overabundant in thorium (and uranium, for the few stars for which measurements are available) with respect to the lanthanides, resulting in negative ages when cosmochronometry is applied.

Key quantities in cosmochronometry are the theoretical production ratios of elements produced by \rp\ events: Th/Eu and U/Eu. 
Over time, radioactive decay decreases these ratios to what is observed today in \rp-enhanced stars.
The extracted ages depend on these ratios as follows:
	\begin{align}
	t &= 46.67~\text{Gyr} \left[\log\epsilon\left(\text{Th/Eu}\right)_0 - \log\epsilon\left(\text{Th/Eu}\right)_{\text{obs}}\right] \label{eqn:theu}\\
	t &= 14.84~\text{Gyr} \left[\log\epsilon\left(\text{U/Eu}\right)_0 - \log\epsilon\left(\text{U/Eu}\right)_{\text{obs}}\right] \label{eqn:ueu}\\
	t &= 21.80~\text{Gyr} \left[\log\epsilon\left(\text{U/Th}\right)_0 - \log\epsilon\left(\text{U/Th}\right)_{\text{obs}}\right] \label{eqn:uth},
	\end{align}
where $\log\epsilon\left(\text{X/Eu}\right)_0$ is the initial production ratio corresponding to the formation of europium and element X at $t=0$, and $\log\epsilon\left(\text{X/Eu}\right)_{\text{obs}}$ is the observed ratio after the radioactive element X has decayed for a time $t$.
The Th/Eu and U/Eu production ratios that have so far been applied to metal-poor stars are largely derived from supernova models \citep[e.g.,][]{wanajo2002,farouqi2010} and fail to account for actinide-boost stars. 
Notably, the thorium-to-uranium (U/Th) ratios produce realistic age estimates in both actinide-boost and non-actinide-boost stars. 

In this work, we study the production of actinides and lanthanides in the low-entropy dynamical ejecta of a NSM. Abundance predictions for this environment depend on the astrophysical conditions assumed, as well as microphysics inputs for the thousands of nuclear species between the valley of stability and the neutron drip line. 
The astrophysical parameters that determine the robustness of the \rp\ are most importantly entropy, dynamical timescale, and electron fraction ($Y_e$), which manifest as temperature and density profiles as a function of time and the initial composition of the nuclear seed material.
The electron fraction is set in part by weak interactions; neutrino emission from the accretion disk (and possibly the hyper-massive neutron star) produced in the merger event can shape the initial composition of accretion disk \citep{surman2006,perego2014,malkus2016} and dynamical outflows \citep{wanajo2014,goriely2015,martin2018}.
Here we focus on the impact of $Y_e$ and select nuclear physics inputs on actinide and lanthanide production.

In Section 2, we describe the nucleosynthesis calculations we use to determine initial production ratios and discuss actinide feeding, including recent calculations of nuclear input that has so far not been used in the context of actinide production. In Section 3, we present a fresh exploration of the production of europium, thorium, and uranium in NSM outflows as a function of neutron excess and investigate the impact of variations in nuclear physics inputs (fission distributions, $\beta$-decay rates, and nuclear masses) on simulated actinide and lanthanide production. In Section 4 we apply our resulting initial production ratios of the cosmochronometer pairs Th/Eu, U/Eu, and U/Th to calculate the age of a recently-discovered actinide boost star and introduce a new method to explain the actinide-boost phenomenon.
We use these age calculations to explore a possible source of the actinide boost.


\section{Actinide Production}


\subsection{Nucleosynthesis Calculations}
\label{sec:calculations}

In the present work, we study \rp\ nucleosynthesis in a NSM trajectory using the Portable Routines for Integrated nucleoSynthesis Modeling (PRISM) nuclear network code (Sprouse et al., \emph{in prep.}). PRISM allows for complete flexibility in nuclear inputs, which we utilize here to investigate the influence of different nuclear physics properties on simulated actinide and lanthanide production. 

We adopt the JINA Reaclib nuclear reaction database \citep{Cyburt+10} for charged-particle and light-nuclei reactions. All relevant \rp\ data are replaced with datasets we construct as self-consistently as possible. Our calculations start with nuclear masses from the Finite Range Droplet Model \citep[FRDM2012;][]{moller2012,moller2016}. We calculate the rates for neutron capture and neutron-induced fission self-consistently with the FRDM2012 masses using the Los Alamos National Laboratory statistical Hauser-Feshbach code \citep{Kawano+16}. Photodissociation rates are calculated using detailed balance. The $\beta$-decay strength functions are from \citet{Moller+18}, and the relative probabilities for $\beta$-decay, $\beta$-delayed neutron emission, and $\beta$-delayed fission are calculated using the QRPA+HF framework \citep{mumpower2016}. All theoretical fission rates use fission barriers from \citet{moller2015}, including the spontaneous fission channel, which is calculated from the relation in \citet{Zagrebaev+11}.

We assume all fission distributions follow a simple symmetric split in which the fissioning nucleus $(Z,A)$, with $Z$ protons, $N$ neutrons, and mass number $A$, splits into two product nuclei $(Z/2,A/2)+(Z/2,A/2)$; no neutron emission is included in this simple treatment. Theoretical $\alpha$-decay rates are found from a Viola-Seaborg relation using $Q_{\alpha}$ values calculated from FRDM2012 masses and parameters fit to known data.
All calculations presented here include evaluated masses and decay rates where known, based on the Atomic Mass Evaluation and {\sc Nubase2016} \citep{audi2017}.
We take care to ensure no theoretical decay rates supplant experimentally established decay data.

For astrophysical conditions, we implement NSM trajectories from a variety of sources and dynamically calculate nuclear reheating, adjusting the temperature of the trajectory accordingly \citep[as in, e.g.,][]{Lippuner+17}. 
We find that $Y_{e}$ and nuclear physics inputs affect the simulated abundances more than the details of reheating \citep{Temis+15}.
Therefore, in this study, we focus on exploring the effects of varying the initial neutron abundance and choices of nuclear physics inputs on the production of the actinides and chronometer pairs. 
We choose a trajectory from the 1.4--1.4 M$_\sun$ NSM simulations of S. Rosswog \citep{rosswog2013,piran2013}, as in \citet{korobkin2012}, with the nominal electron fraction of $Y_e=0.035$. 
We begin calculations at a temperature of 10~GK with seed distributions in nuclear statistical equilibrium (NSE) calculated with the SFHo/FRDM model from \citet{steiner2013}.
All final abundances shown are at 1~Gyr after the start of the \rp\ event.


\subsection{Actinide Feeding}

\begin{figure}[t]
 \centerline{\includegraphics[width=\columnwidth]{{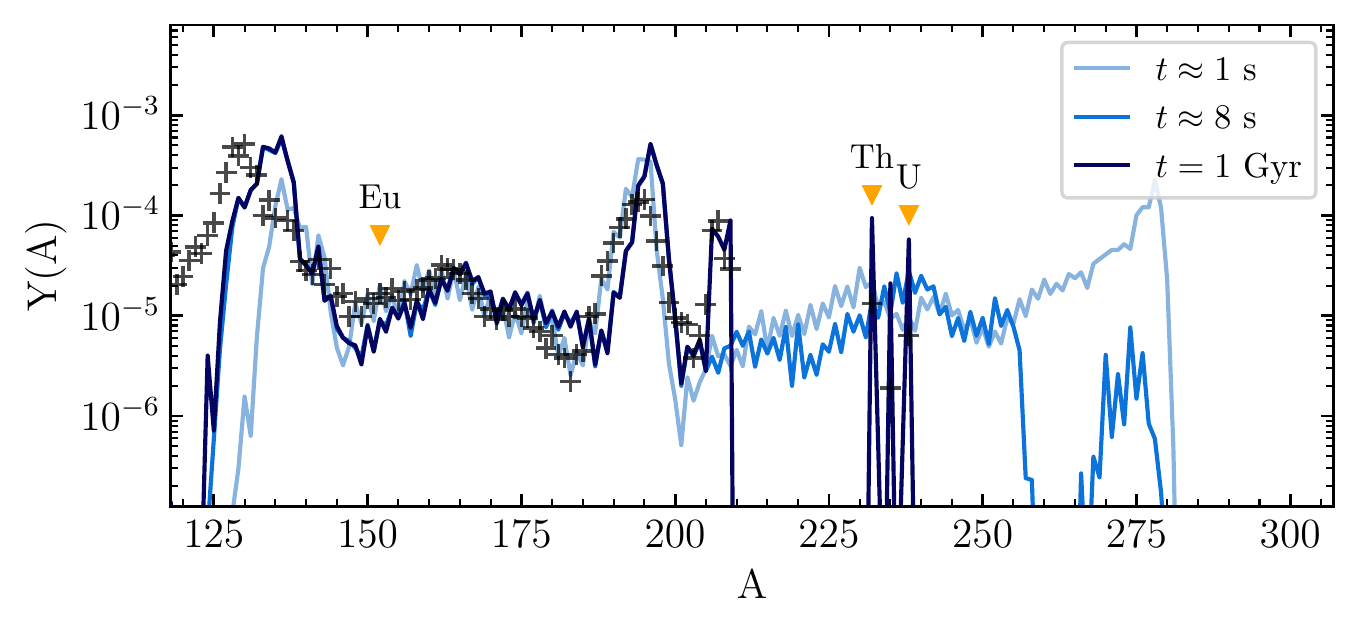}}}
 \caption{\label{fig:times_frdm2012} Isotopic abundance patterns for the baseline simulation at \rp\ freezeout ($t\approx 1$ s), when the $\beta$-delayed fission dominates over neutron-induced fission ($t\approx 8$ s), and the final time ($t=1$ Gyr). The isotopes $^{232}$Th, $^{238}$U, and the region of Eu ($A=151,153$) are indicated.}
\end{figure}

Metal-poor, \rp-enhanced stars are sufficiently old that the only isotopes contributing to the total actinide abundance are $^{232}{\rm Th}$ and $^{238}{\rm U}$.
These two nuclei are primarily populated by $\beta$-decay and through $\alpha$-decay chains of heavier nuclei. Here we investigate their population in the FRDM2012 (``baseline") simulation with starting electron fraction of $Y_e=0.035$. 

The net amount of material available to eventually feed $^{232}{\rm Th}$ and $^{238}{\rm U}$ is determined by the nuclear flow of material into the actinide region of the $N$-$Z$ plane.
The $N=126$ shell closure, in particular, moderates the flow into this region.
The flow of material out of this region is determined by fission and, eventually, $\alpha$ decay.
The evolution of the abundances as material populates the actinide region is shown in Figure \ref{fig:times_frdm2012} at key times in the \rp\ for the baseline simulation.

At early times, sufficient nuclear reheating occurs such that the nuclear flow proceeds largely in $(n,\gamma)$-$(\gamma,n)$ equilibrium; nuclear masses therefore set the location of the \rp\ path, and the flow through the $N=82$, 126, and 184 closed shells is regulated by the $\beta$-decay lifetimes of the waiting points at each shell closure. Here the \rp\ path is terminated in $A$ by neutron-induced fission just above the predicted $N=184$ shell closure. At these early stages of the \rp, the abundances in the actinide region and above are shaped by the strengths of the predicted $N=126$ and 184 shell closures, the $\beta$-decay lifetimes in the $N=126$ region and above, and by the fission barrier heights that determine where the \rp\ path terminates in $A$.

As the neutron flux decreases significantly, the system falls out of $(n,\gamma)$-$(\gamma,n)$ equilibrium.
The abundance pattern for the baseline simulation at this ``freezeout time" is shown by the light blue line (labeled ``$t=1$ s") in Figure \ref{fig:times_frdm2012}.
At this time, a $N=184$ closed shell peak is evident at $A\sim 275$, and the abundances drop sharply at higher $A$ due to the onset of neutron-induced fission above the closed shell.

As the neutron abundance continues to drop precipitously, the rates for neutron-induced fission also decline, and eventually $\beta$-delayed fission takes over as the dominant fission channel \citep{thielemann1983,panov2004,petermann2012,mumpower2018}.
The abundances at this time are indicated by the medium blue line in Figure \ref{fig:times_frdm2012} (labeled ``$t\approx 8$ s"). Between the first two times in Figure~\ref{fig:times_frdm2012}, about 78\% of the mass above $A=230$ leaves the heavy region as fission transfers material to lower mass numbers.
By the end of the simulation, at $t=1$ Gyr, approximately 10\% of the original $A>230$ mass finds its way into either $^{232}$Th or $^{238}$U. 
The remaining $A>230$ mass (1) populates other short-lived actinides, (2) continues $\alpha$-decay to populate lead and bismuth, or (3) simply fissions.
The abundance pattern at this final time is shown by the dark blue line in Figure \ref{fig:times_frdm2012}.

\begin{figure}[t]
 \centerline{\includegraphics[width=\columnwidth]{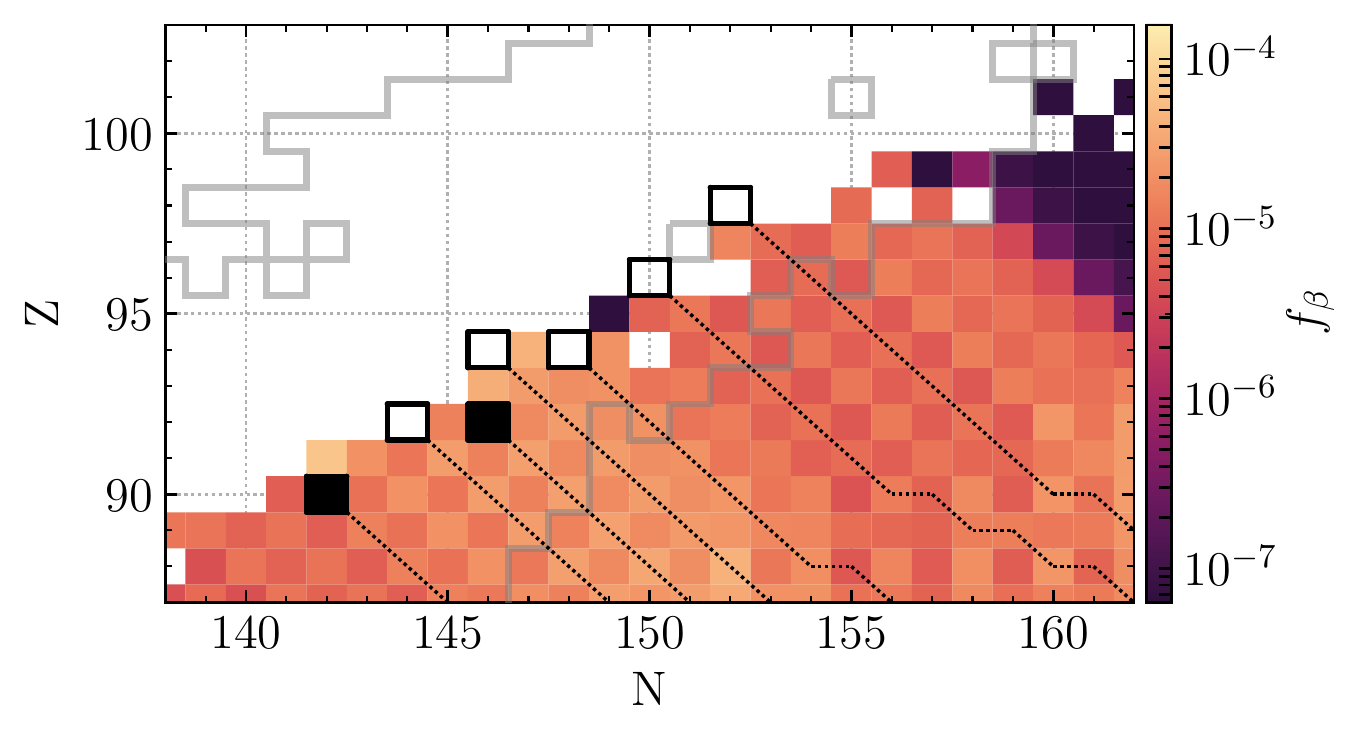}} \vspace{-2mm}
 \centerline{\includegraphics[width=\columnwidth]{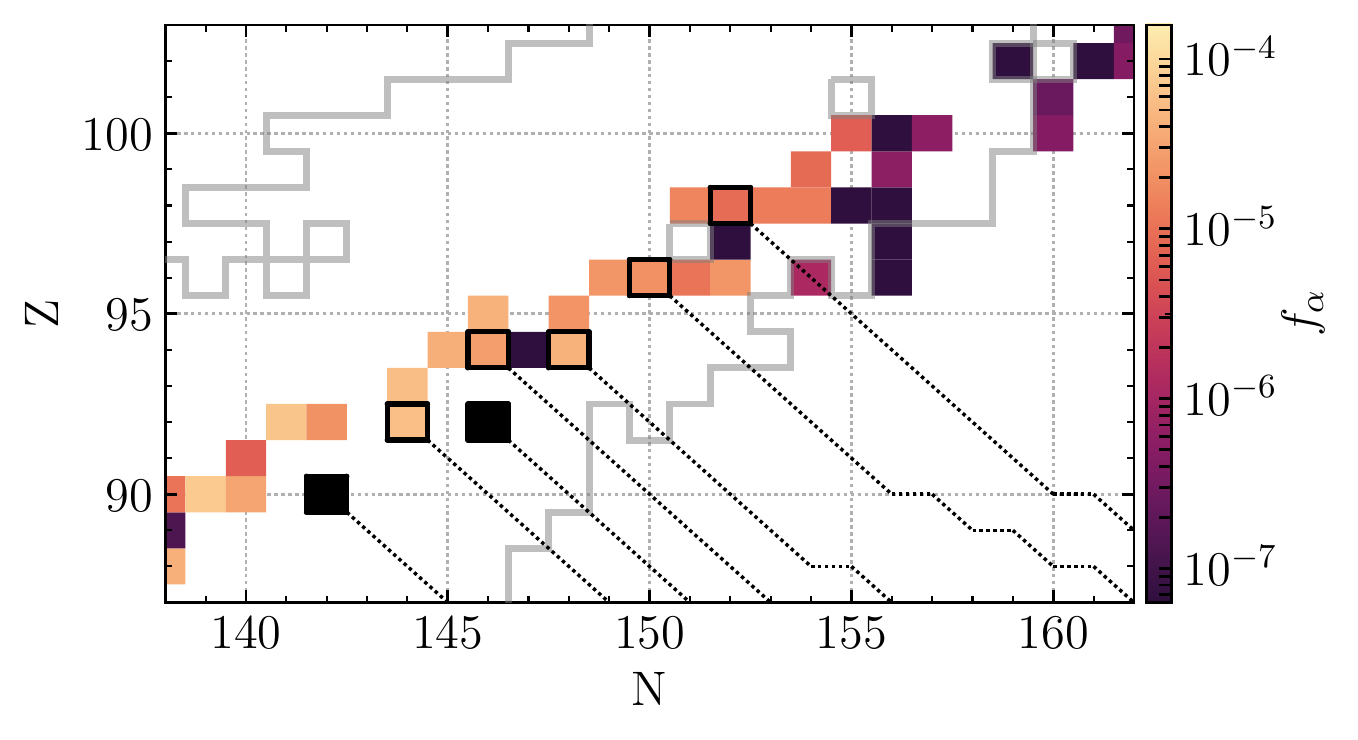}} 
 \caption{\label{fig:flow_frdm2012} Integrated flows for the base calculation at $Y_e=0.035$ for $\beta$-decay (top) and $\alpha$-decay (bottom). Filled boxes indicate $^{232}$Th and $^{238}$U. Bold outlined boxes denote nuclei that $\alpha$-decay into $^{232}$Th or $^{238}$U, and dashed diagonal lines are the most probable $\beta$-decay paths feeding those nuclei. The region outlined in gray denotes nuclei which are included in {\sc Nubase2016}.}
\end{figure}

The late-time population of the actinides can be followed in detail using integrated reaction flows. Figure~\ref{fig:flow_frdm2012} shows the integrated $\beta$- and $\alpha$-decay flows for the baseline calculation.
For discrete timesteps, the integrated reaction flow $f_x(Z,A)$ of nucleus $(Z,A)$ is expressed as
	\begin{equation}
	f_x(Z,A) = \sum_i \lambda_{x,i}(Z,A)\, Y_i(Z,A)\, \left(t_{i+1} - t_i\right), 
	\label{eqn:flow}
	\end{equation}
where $i$ is the timestep, $t_i$ the time at timestep $i$, $\lambda_{x,i}(Z,A)$ the rate for reaction $x$ for nucleus $(Z,A)$ at time $t_i$, and $Y_i(Z,A)$ the abundance of nucleus $(Z,A)$ at time $t_i$.
The reaction flow at time $t_i$ is $\lambda_{x,i}(Z,A)\, Y_i(Z,A)$.

Five nuclei are primarily responsible for the population of observable thorium and uranium by direct $\alpha$-decay chains; these actinide ``feeders" are indicated by bold boxes in Figure~\ref{fig:flow_frdm2012}.
The primary $\alpha$-decay feeders of $^{232}$Th are $^{236}$U and $^{240}$Pu. Population of nuclei higher up this $\alpha$ chain, $^{244}$Cm and $^{248}$Cf, are effectively blocked by long-lived $\alpha$-emitters $^{244}$Pu and $^{248}$Cm (80 and 0.35 Myr, respectively). $^{238}$U is fed via the $\alpha$ decays of $^{242}$Pu, $^{246}$Cm, and $^{250}$Cf. The $\beta$-feeding of the next-higher nucleus on this $\alpha$ chain, $^{254}$Fm, is blocked by the spontaneous fission of $^{254}$Cf \citep{zhu+2018}, and its $\alpha$-feeding is prevented by the spontaneous fission of $^{258}$No.

These $\alpha$-chain feeders are populated by $\beta$ decay.
Close to stability, $\beta$-feeders can be traced directly to their parent nuclei.
However, farther from stability, the probability of $\beta$-delayed neutron-emission increases, and many parent isotopes may decay to the same daughter nucleus.
Consequently, many nuclei far from stability contribute to the total abundance of the $\alpha$-feeders through complex $\beta$-decay pathways.
The $\beta$-feeding pathways following the highest $\beta$-delayed neutron-emission flows are indicated by the dotted lines in Figure~\ref{fig:flow_frdm2012}.

The solid gray line in Figure~\ref{fig:flow_frdm2012} indicates the extent of evaluated decay data from {\sc Nubase2016}. Included in this database are the half-lives of all of the primary $\alpha$-feeders, as well as the spontaneous fission branchings that directly impact the $^{232}$Th and $^{238}$U feedings.
Thus the theoretical $\alpha$-decay and spontaneous fission rates we implement outside of this region have no substantive quantitative impact on the final simulated abundances of $^{232}$Th and $^{238}$U, since many nuclei surrounding $^{232}$Th and $^{238}$U are well-studied experimentally.
We confirmed this negligible effect by rerunning the baseline calculation with a variety of choices for theoretical $\alpha$-decay and spontaneous fission rates found in literature \citep[e.g.,][]{Zagrebaev+11,swiatecki1955,petermann2012}, including removing them completely from the calculation. We find that the sensitivity of the final abundances to these choices of theoretical spontaneous fission and $\alpha$-decay rates is less than 0.1\%.

Instead, the major theoretical nuclear data sets that affect our calculations of actinide feeding include fission fragment distributions, $\beta$-decay rates and branchings, and nuclear masses. In the next section we describe the production of thorium and uranium relative to europium as a function of the neutron richness of the astrophysical conditions for distinct choices of the nuclear physics.


\section{Production of Eu, Th, and U}
\label{sec:production}


\subsection{Baseline Calculation: FRDM2012}
\label{sec:baseline}

\begin{figure}[t]
 \centerline{\includegraphics[width=0.95\columnwidth]{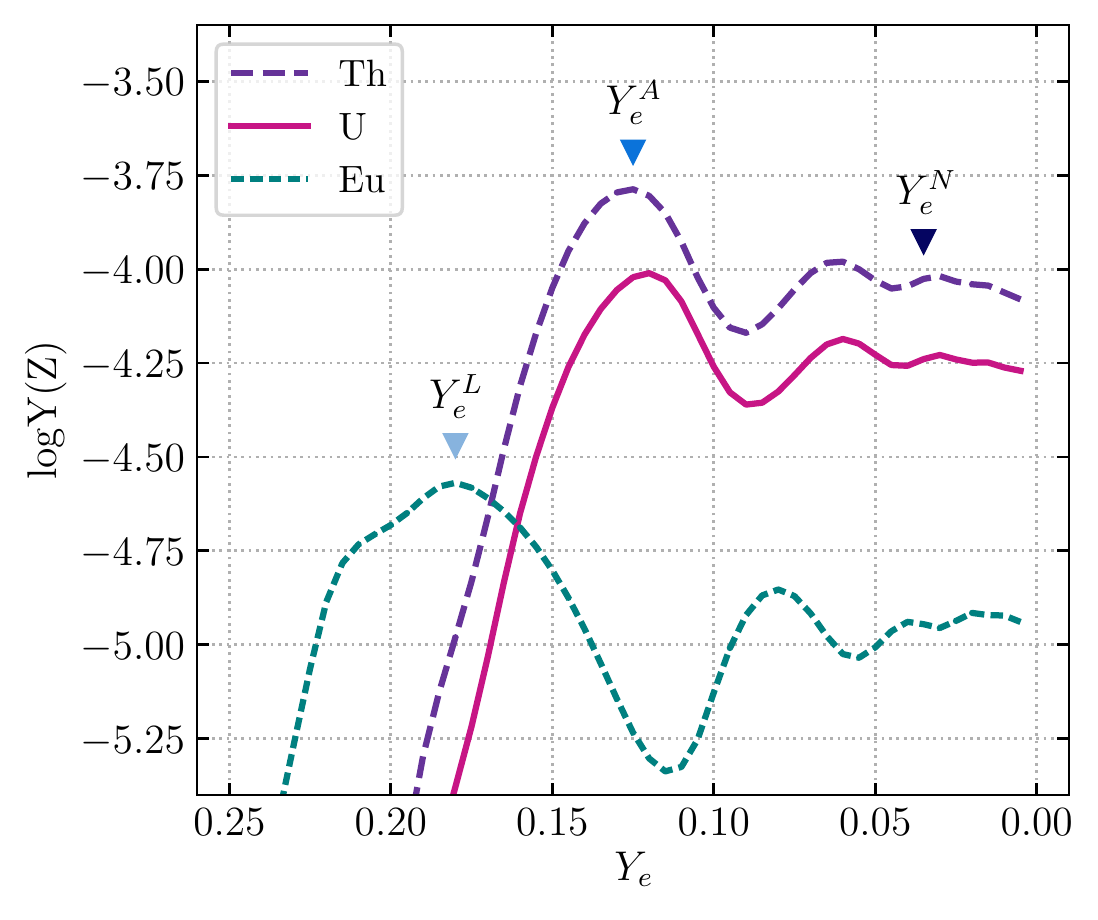}}
 \caption{\label{fig:ye_frdm2012} Europium, thorium, and uranium production as a function of initial $Y_e$ using 50/50 fission fragment distributions and the FRDM2012 mass model. The $Y_e$ is shown on a reversed scale to reflect increasing neutron richness.}
\end{figure}

\begin{figure}[t]
 \centerline{\includegraphics[width=0.85\columnwidth]{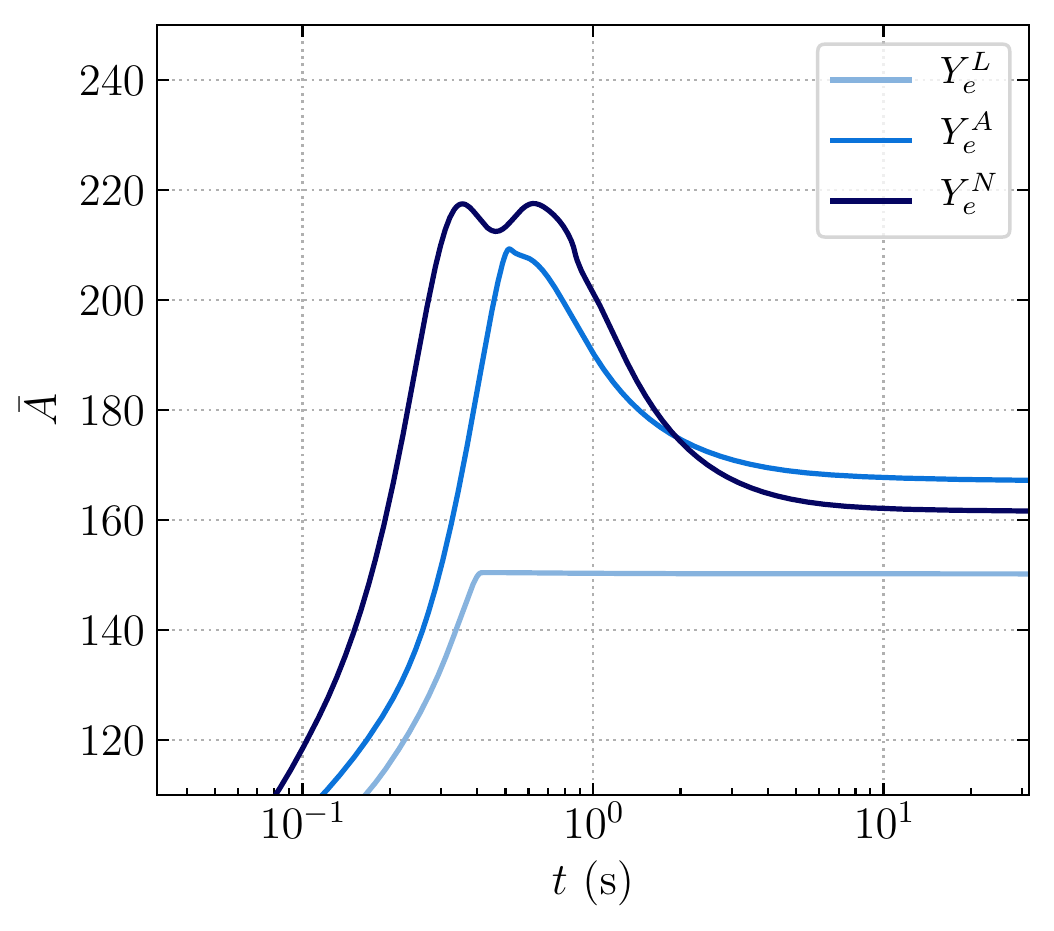}}
 \caption{\label{fig:avgA_frdm2012} Average mass as a function of time for $Y_e^L$, $Y_e^A$, and $Y_e^N$ for the baseline case, using symmetric fission fragment distributions and the FRDM2012 mass model. }
\end{figure}

\begin{figure}[t]
 \centerline{\includegraphics[width=\columnwidth]{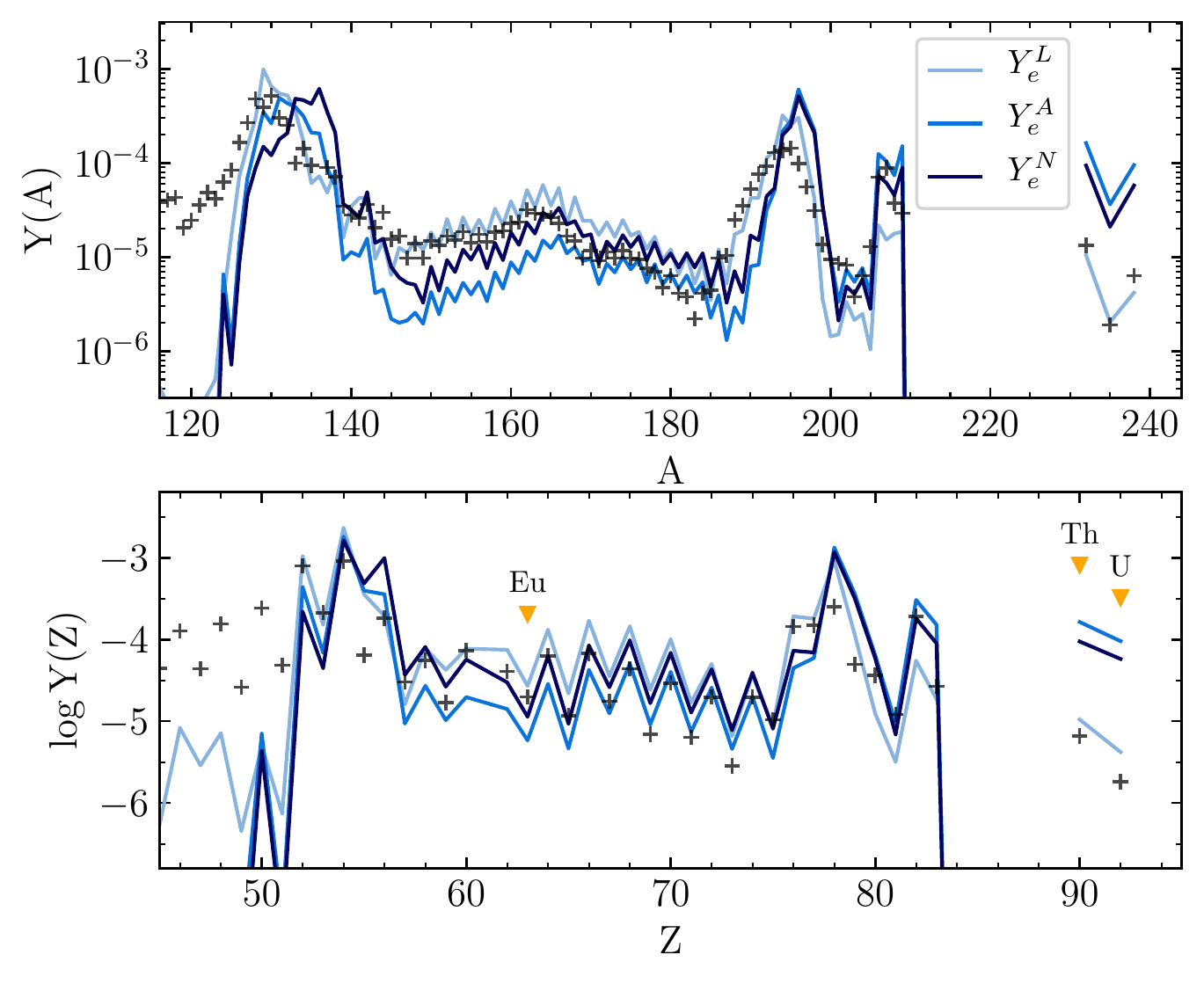}}
 \caption{\label{fig:ab_frdm2012} Final isotopic (top) and elemental (bottom) abundance patterns for the baseline case corresponding to the three initial $Y_e$ values where lanthanides are maximized ($Y_e^L$), actinides are maximized ($Y_e^A$), and the nominal value ($Y_e^N$). Scaled-Solar \rp\ abundances (``+") are from \citet{sneden2008}. The elements Eu, Th, and U are indicated in the bottom panel for clarity.}
\end{figure}

For our baseline calculation set, we take FRDM2012 nuclear masses and nuclear data as described in Section 2.1 and vary the initial electron fraction ($Y_e$) of the baseline astrophysical trajectory between 0.005 and 0.250 in equal intervals of 0.005. We run a full \rp\ nucleosynthesis calculation for each starting $Y_e$.
Figure~\ref{fig:ye_frdm2012} shows the production of total europium\footnote{Only $^{151}$Eu and $^{153}$Eu are produced by the \rp\ and are stable on these timescales.}, $^{232}$Th, and $^{238}$U as a function of initial $Y_e$ for the 50 \rp\ simulations.
We hereafter refer to the abundances of these nuclei as generally ``thorium" (or ``Th") and ``uranium" (or ``U").
For this analysis, we focus on three values of $Y_e$ in particular:
	\begin{itemize}
	\item $Y_e^L$: When Eu (a \emph{L}anthanide) reaches the first local maximum as $Y_e$ decreases
	\item $Y_e^A$: When Th (an \emph{A}ctinide) is maximized
	\item $Y_e^N$: The \emph{N}ominal value of 0.035 from \citet{korobkin2012}
	\end{itemize}
These critical $Y_e$ values for the FRDM2012 (baseline) case are denoted in Figure~\ref{fig:ye_frdm2012} and analyzed in detail in Figures~\ref{fig:avgA_frdm2012} and \ref{fig:ab_frdm2012}. Figure~\ref{fig:avgA_frdm2012} shows the average mass number ($\bar{A}$) as a function of time for simulations with starting $Y_e$ values of $Y_e^L$, $Y_e^A$, and $Y_e^N$, and Figure~\ref{fig:ab_frdm2012} presents their final isotopic and elemental abundance patterns.

As can be seen in Figure \ref{fig:ye_frdm2012} at the highest $Y_{e}$ considered (0.25), the \rp\ is insufficiently neutron-rich to proceed past the $N=82$ shell closure, and $A>140$ nuclei are not significantly populated. 
Simulations with increasing neutron-richness, $0.185<Y_e<0.25$, produce increasing amounts of the lanthanides, with Eu production reaching a maximum at $Y_e=0.185=Y_e^L$.
The $Y_e^L$ simulation is sufficiently neutron-rich to produce the $N=126$ closed-shell peak but not neutron-rich enough to move much material beyond it. 
As a result, the majority of mass in the network is within and between the $N=82$ and $N=126$ closed-shell peaks. 
The evolution of average mass number $\bar{A}$ for this simulation is indicated by the light blue line of Figure~\ref{fig:avgA_frdm2012}. 
$\bar{A}$ initially increases as the \rp\ path is populated and the nuclear flow proceeds to higher mass numbers. 
Once the free neutrons are exhausted and the material begins to decay toward stability, at around $t\sim 0.4$ s, $\bar{A}$ flattens out to its final value of $\bar{A} \sim \mbox{150}$. 
The final abundance pattern in Figure~\ref{fig:ab_frdm2012} includes robust $A\sim 130$ and rare-earth ($A\sim 160$) peaks, with a deficit of material beyond $A\sim 200$.

In simulations with $Y_e<Y_e^L$, neutron capture continues past the $N=126$ closed-shell peak and populates the actinides more effectively. 
As the material moves up to higher $A$, the lanthanides are correspondingly depopulated. 
At $Y_e=0.125=Y_e^A$, as indicated in Figure~\ref{fig:ye_frdm2012}, the Th abundance reaches a maximum and Eu a local minimum. 
The evolution of $\bar{A}$ for the simulation with $Y_e^A$ is indicated by the medium blue line of Figure~\ref{fig:avgA_frdm2012}. 
Here most of the mass of the simulation moves beyond the $N=82$ peak and $\bar{A}$ exceeds 200 at freezeout, which occurs around $t\sim 0.5$ s in this case. 
After freezeout, the drop in $\bar{A}$ is due to the depopulation of the $A>230$ region by fission and $\alpha$ decay, as described in Section 2.2. 
The fission products are deposited in the $A\sim 130$ region, and insufficient neutrons remain for the products to capture out of this region.
This effect produces a final abundance pattern with strong closed-shell peaks and under-abundant lanthanides, as shown by the medium blue line of Figure~\ref{fig:ab_frdm2012}.

Simulations with $Y_e<Y_e^A$ have increased availability of neutrons to both induce fission---removing material from the actinides---and to allow fission products to capture into the rare-earth region.
Correspondingly, in simulations with $0.08<Y_{e}<0.125$, the europium abundance again increases and the actinides decrease with decreasing $Y_e$, as seen in Figure~\ref{fig:ye_frdm2012}.
A second local europium-minimum/actinide-maximum is found at $Y_{e}=0.06$, which occurs when sufficient neutrons are present for fission products to capture past the $N=126$ shell closure and back into the actinide region. 
At this point the integrated fission flow summed over the nuclear chart, $\sum_{Z,A} f_x(Z,A)$, roughly equals the total abundance of the network $\sum_{Z,A} Y(Z,A)$, suggesting the onset of fission recycling.

Fission recycling becomes increasingly robust for simulations with decreasing $Y_{e}<0.06$. At $Y_e^N$, the evolution of $\bar{A}$ shows evidence of multiple fission cycles, shown by the oscillatory patten in Figure~\ref{fig:avgA_frdm2012} at $\bar A>200$, which ends at freezeout ($t\sim 0.8$ s). 
Material is deposited into the second peak and recycled several times, allowing the abundance patterns in Figure~\ref{fig:ab_frdm2012}---in particular the europium abundance---to stabilize between the two extremes of the $Y_e^L$ and $Y_e^A$ cases \citep{beun2008,Temis+15}.


\subsection{Dependence on fission fragment distribution: \\Kodama and Takahashi}
\label{sec:kodama}

\begin{figure}[t]
 \centerline{\includegraphics[width=0.95\columnwidth]{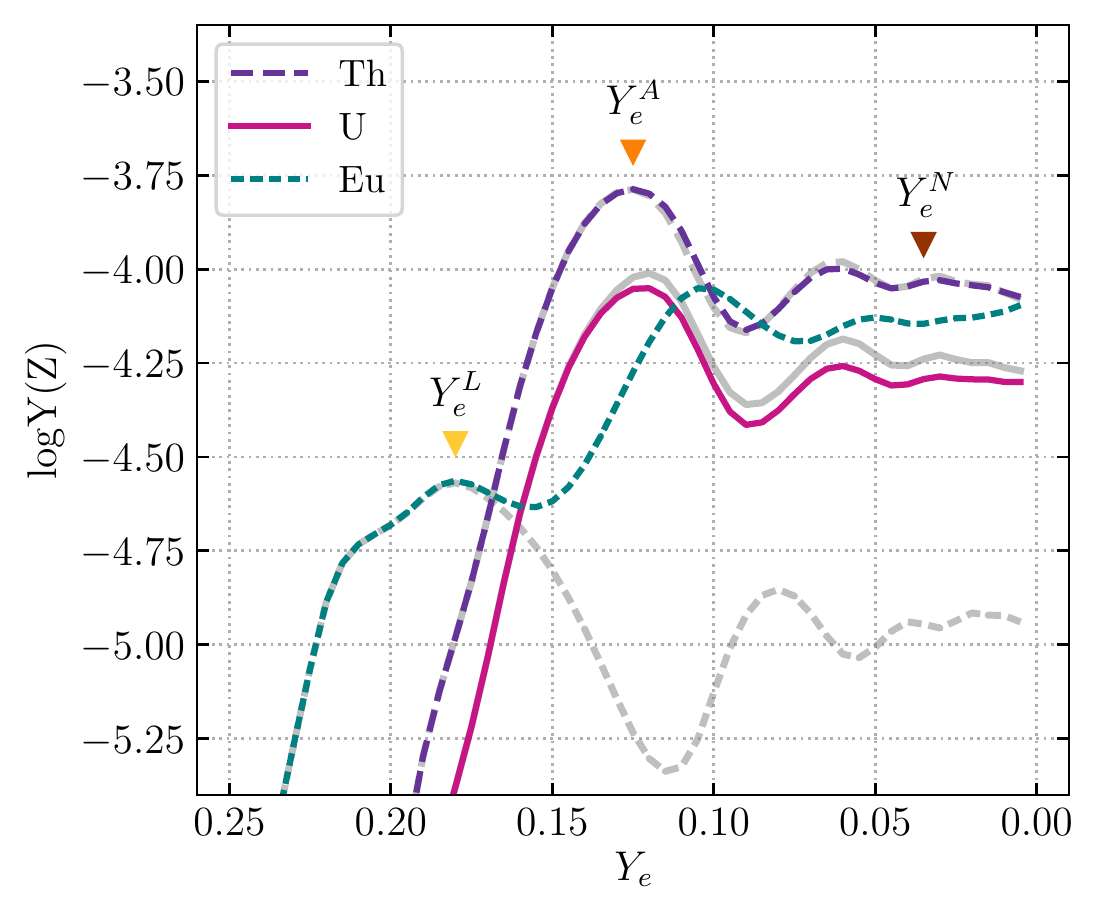}}
 \caption{\label{fig:ye_kodama} Europium, thorium, and uranium production as a function of initial $Y_e$ using \citet{kodama1975} fission distributions and the FRDM2012 mass model. The baseline case abundances of Figure~\ref{fig:ye_frdm2012} are plotted in gray for comparison.}
\end{figure}

At low $Y_{e}$, the shape of the \rp\ pattern is dominated by fission recycling and is therefore dependent on the fission properties of neutron-rich nuclei. The fission barriers, rates, and product distributions of these nuclei are poorly known; little experimental data are available, and theoretical estimates vary widely, leading to significant differences in the \rp\ pattern \citep{eichler2015,Cote+18}. Here we examine the impact of fission product yields on lanthanide and actinide production by repeating the simulations of Section 3.1 while replacing the 50/50 simple split fission fragment distributions with the double-Gaussian fission distributions of \citet{kodama1975} (hereafter ``\kt").

\begin{figure}[t]
 \centerline{\includegraphics[width=0.85\columnwidth]{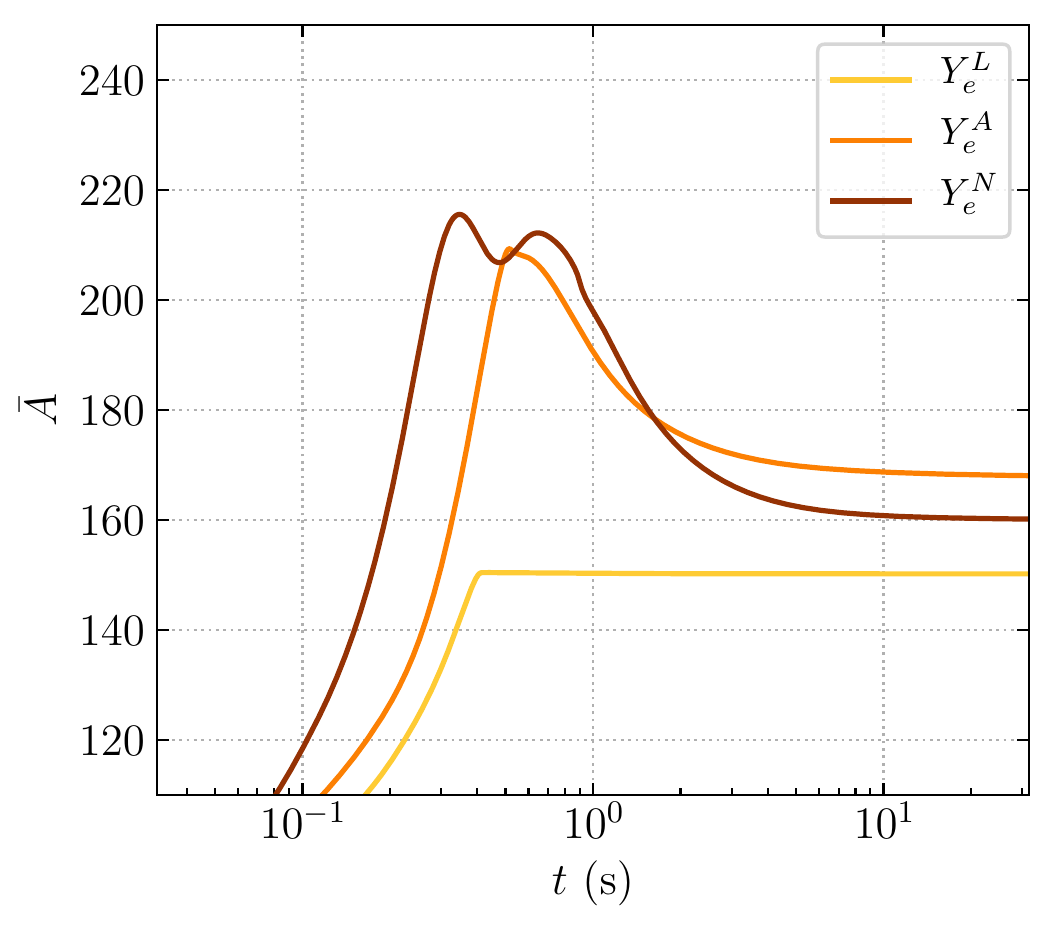}}
 \caption{\label{fig:avgA_kodama} Average mass as a function of time for $Y_e^L$, $Y_e^A$, and $Y_e^N$ for the \kt\ case.}
\end{figure}

\begin{figure}[t]
 \centerline{\includegraphics[width=\columnwidth]{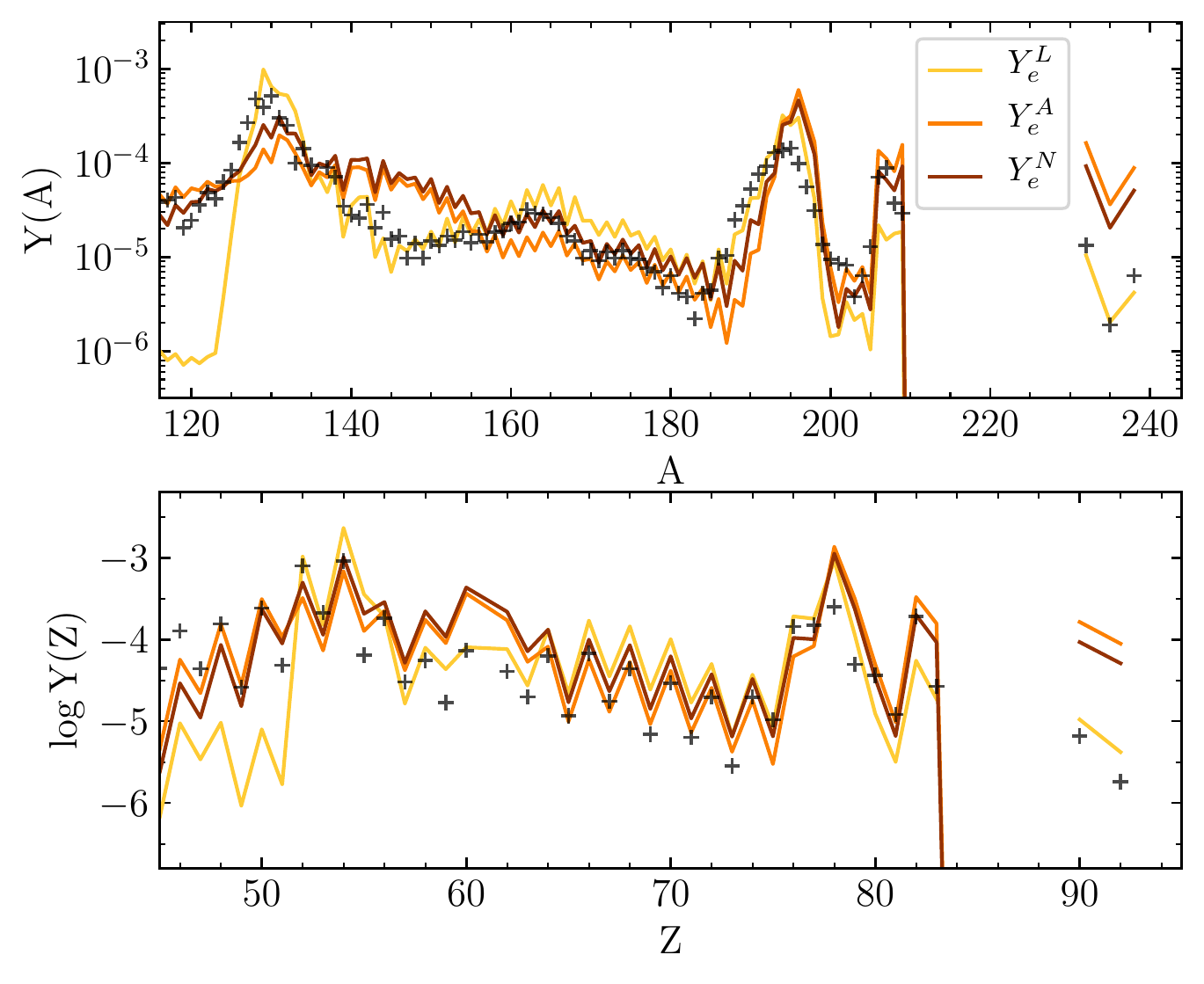}}
 \caption{\label{fig:ab_kodama} Final isotopic (top) and elemental (bottom) abundance patterns for the \kt\ case corresponding to the three initial $Y_e$ values where lanthanides are locally maximized ($Y_e^L$), actinides are maximized ($Y_e^A$), and the nominal value ($Y_e^N$). Scaled-Solar \rp\ abundances (``+") are from \citet{sneden2008}.}
\end{figure}

Figure~\ref{fig:ye_kodama} shows the effect of using the \kt\ fission fragment distribution on the europium and actinide abundances compared to the simple 50/50 split.
As expected, Th and U production is similar in both cases.
The small differences that do arise between the abundances of these nuclei are due to the width of the \kt\ fission distribution, which allows more material to neutron capture up into the fissioning region.
This width results in slightly more neutron-induced fission than the baseline case, leading to slightly smaller $\bar{A}$ during fission recycling, as shown in Figure~\ref{fig:avgA_kodama} for the $Y_e^N$ simulation, and slightly lower actinide production for $Y_e<Y_e^A$ as indicated in Figure~\ref{fig:ye_kodama}.

In contrast to thorium and uranium, europium exists at an atomic mass that can be significantly affected by fission fragment distribution.
The \kt\ fission distribution probabilities are modeled by very broad Gaussians, and our simulations using these yields show fission product deposition in a wide region around the $A\sim 130$ peak, including in the lanthanide region.
The additional direct deposition of fission products significantly increases the final Eu abundances for all simulations with an appreciable amount of fission, as shown in Figure~\ref{fig:ye_kodama}.
The resulting final abundance patterns at $Y_e^L$, $Y_e^A$, and $Y_e^N$ are shown in Figure~\ref{fig:ab_kodama}.
Even after a single episode of fission cycling ($Y_e^A$), the $A\sim 145$ region is completely reshaped by the distribution of fission products.
Although the actinides are relatively unaffected, fission fragment distributions may impact the lanthanides directly, reshaping the Th/Eu and U/Eu production ratios.


\subsection{Dependence on $\beta$-decay rates: Marketin}
\label{sec:marketin}

\begin{figure}[t]
 \centerline{\includegraphics[width=0.95\columnwidth]{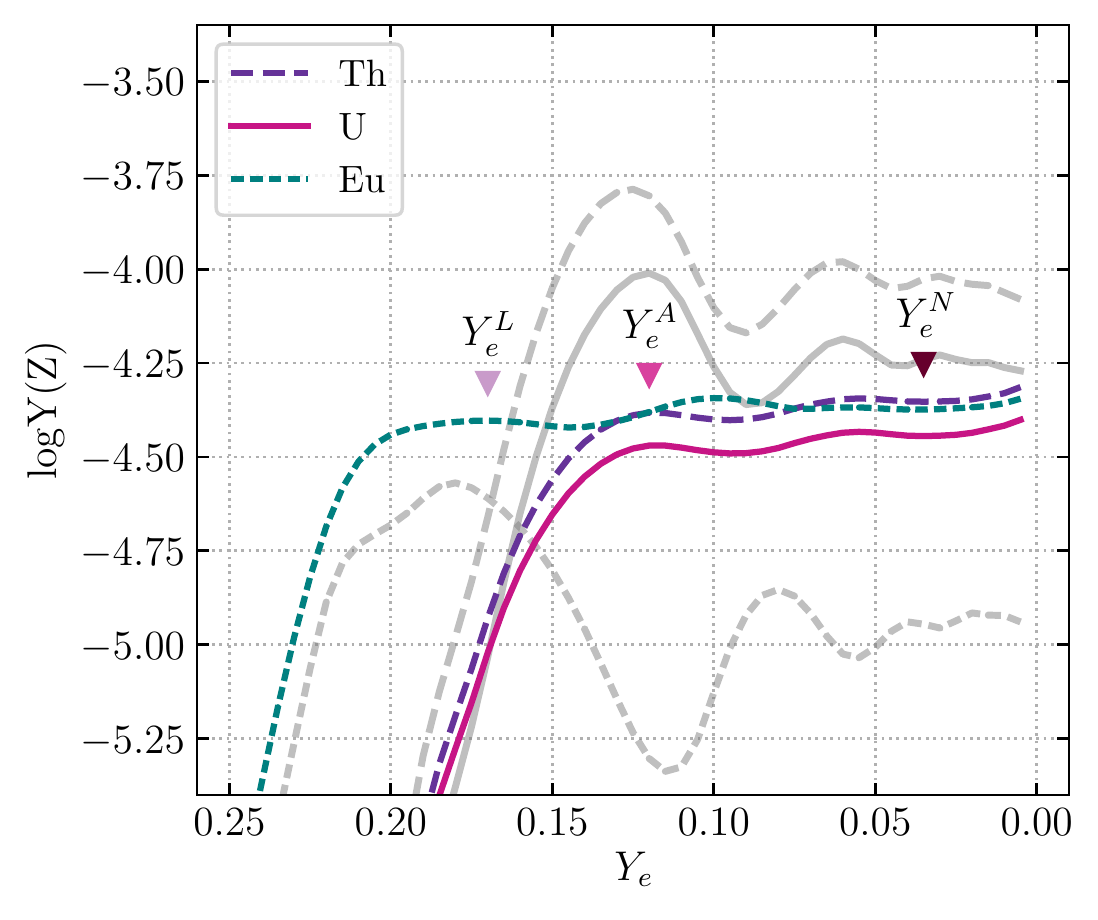}}
 \caption{\label{fig:ye_marketin} Europium, thorium, and uranium production as a function of initial $Y_e$ using 50/50 fission fragment distributions, \citet{marketin2016} $\beta$-decay rates, and the FRDM2012 mass model. The baseline case abundances of Figure~\ref{fig:ye_frdm2012} are plotted in gray for comparison.}
\end{figure}

The $\beta$-decay half-lives of neutron-rich nuclei influence all phases of the \rp. At early times they set the timescale for nuclear flow to high $A$, the rate of nuclear reheating, and the relative abundances along the \rp\ path. During freezeout, the final abundances are determined from a competition between $\beta$-decay and all other available reaction channels.
To study the impact of theoretical $\beta$-decay rates on lanthanide and actinide production, we start with our baseline case and use $\beta$-decay rates from \citet{marketin2016} in place of \citet{Moller+18}.
Reheating is recalculated self-consistently with the updated rates.

Figure~\ref{fig:ye_marketin} shows the resulting abundance evolution as a function of $Y_e$.
The same cyclic pattern as the baseline case can still be identified, however, with the \citet{marketin2016} rates the Eu production is larger, the U and Th abundances are lower, and the ratio of U to Th is higher.
These differences are driven by two distinct regions of disparity between the \citet{marketin2016} and the baseline \citet{Moller+18} $\beta$-decay rates.

\begin{figure}[t]
 \centerline{\includegraphics[width=\columnwidth]{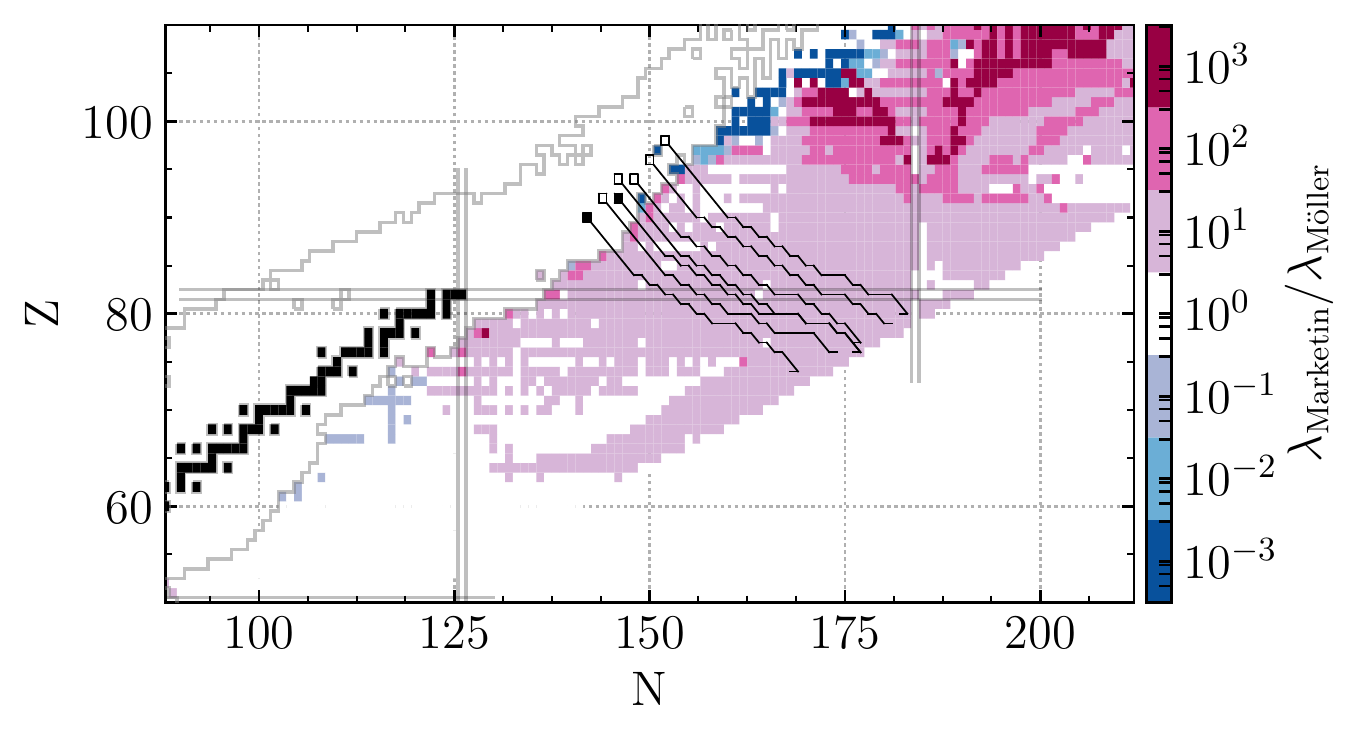}}
 \centerline{\includegraphics[width=\columnwidth]{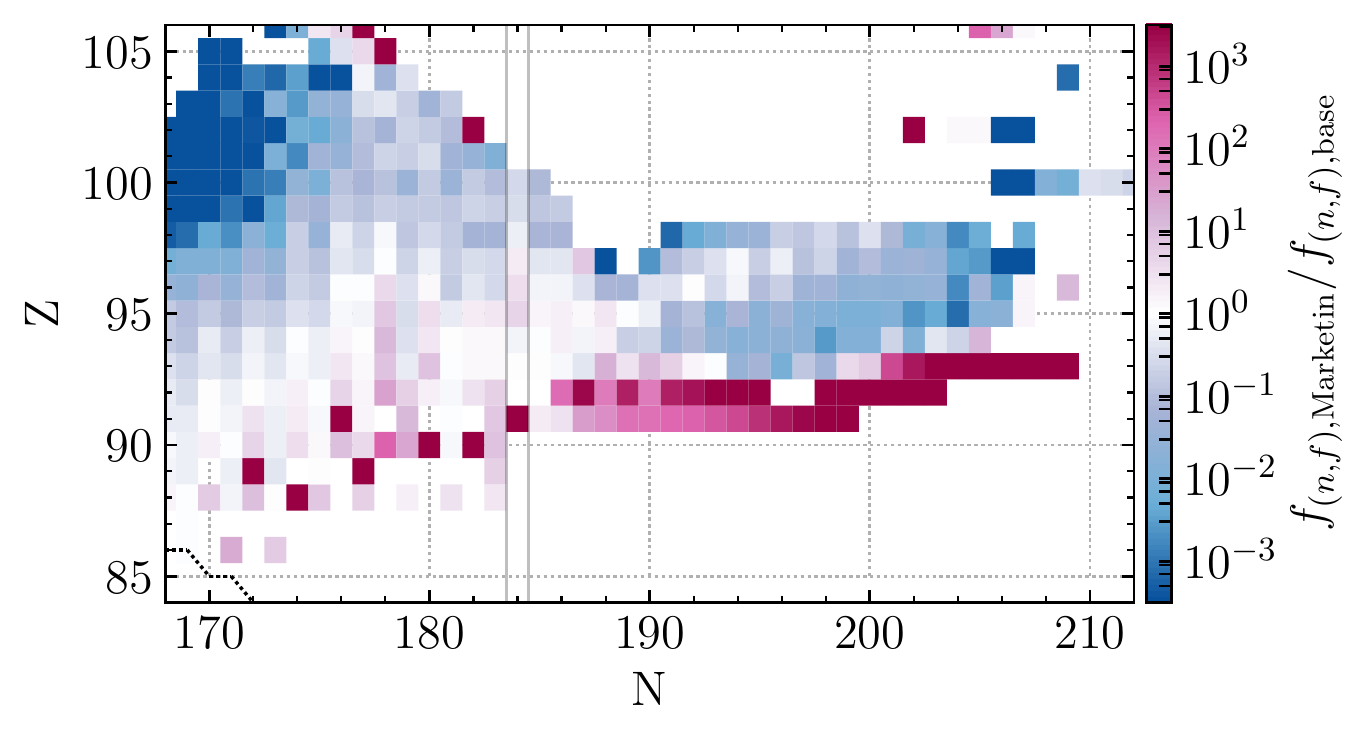}} 
 \caption{\label{fig:beta_compare} Top: comparison of the $\beta$-decay rates between QRPA+HF and \citet{marketin2016}. Bottom: integrated neutron-induced fission flow comparison of the \citet{marketin2016} case and the baseline at $Y_e=0.035$.}
\end{figure}

The first relevant difference is in the $A=130$ peak region.
In particular, the \citet{Moller+18} $\beta$-decay rates along the $Z=48$ isotopic chain near the doubly-magic shell closure are approximately four times slower than the \citet{marketin2016} rates. Differences greater than a factor of two are also present for the $Z=44$ and $Z=46$ isotopic chains. When applied to \rp\ simulations, the faster \citet{marketin2016} rates of these isotopic chains result in less material piling up in the $A=130$ region and more material filling the rare-earth region, producing the increased amount of Eu in Figure~\ref{fig:ye_marketin} relative to the baseline.

\begin{figure}[t]
 \centerline{\includegraphics[width=0.85\columnwidth]{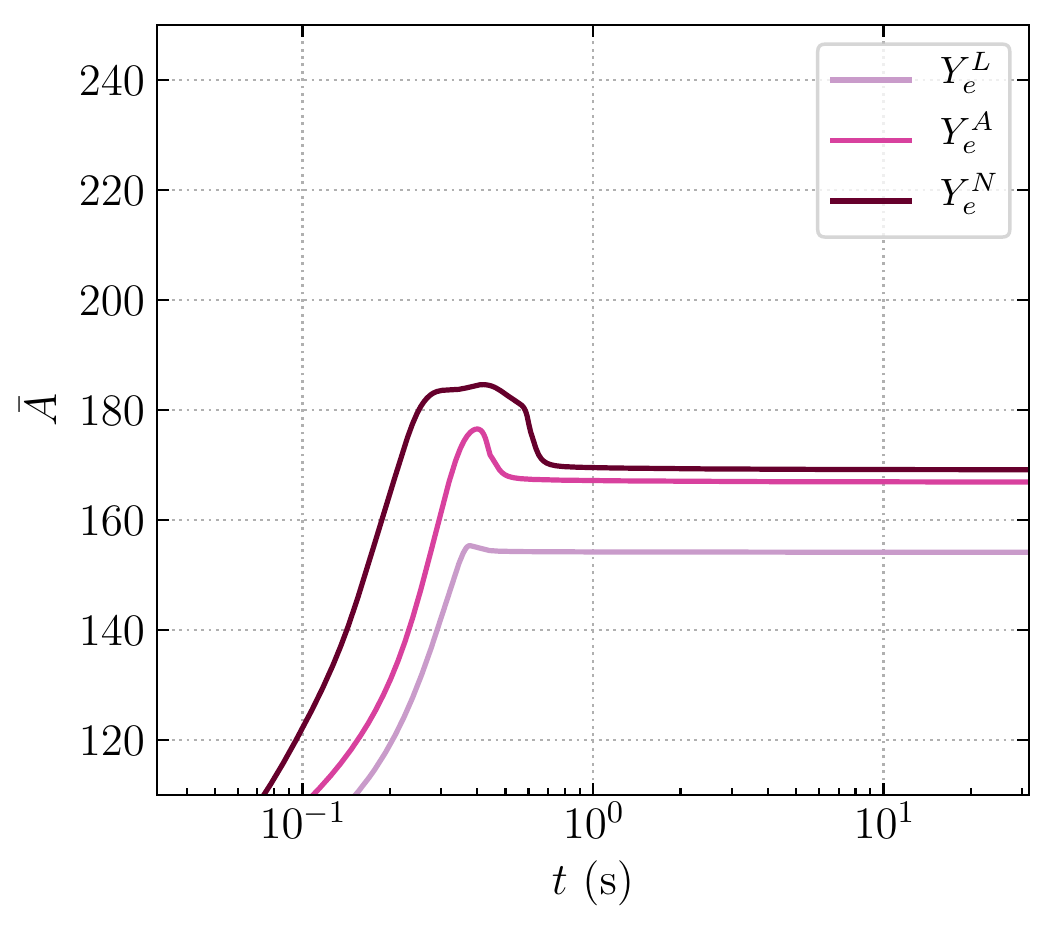}}
 \caption{\label{fig:avgA_marketin} Average mass as a function of time for $Y_e^L$, $Y_e^A$, and $Y_e^N$for the Marketin case.}
\end{figure}

\begin{figure}[t]
 \centerline{\includegraphics[width=\columnwidth]{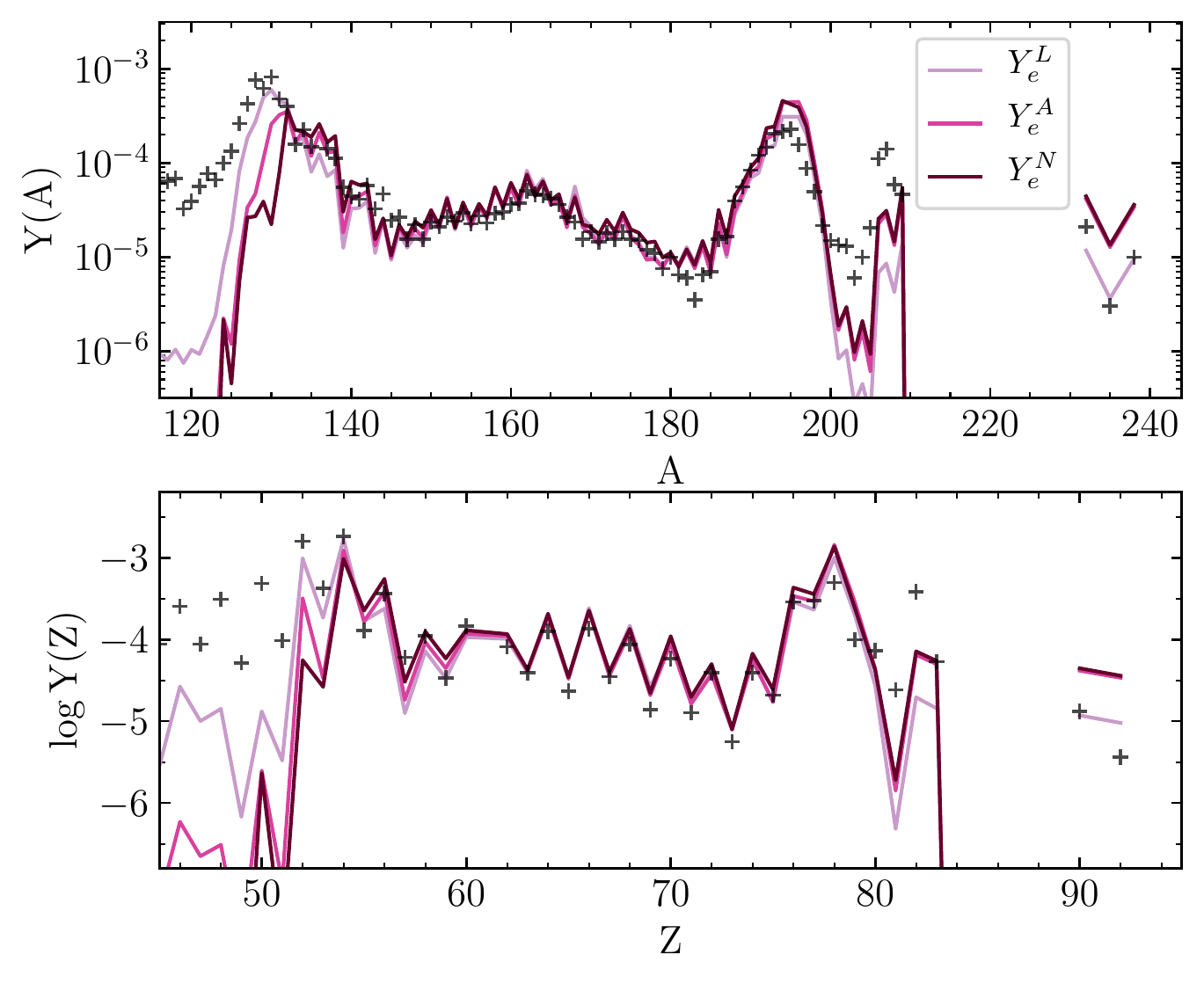}}
 \caption{\label{fig:ab_marketin} Final isotopic (top) and elemental (bottom) abundance patterns for the Marketin case corresponding to the three initial $Y_e$ values where lanthanides are locally maximized ($Y_e^L$), actinides are locally maximized ($Y_e^A$), and the nominal value ($Y_e^N$). Scaled-Solar \rp\ abundances (``+") are from \citet{sneden2008}.}
\end{figure}

The second and larger region of disparity is at high $A$. The \citet{marketin2016} $\beta$-decay rates range between two and several thousand times faster than the \citet{Moller+18} rates above $A\sim190$, as shown in Figure~\ref{fig:beta_compare}. 
The faster $\beta$-decay rates of \citet{marketin2016} beyond the third \rp\ peak allow material to pass through the predicted $N=184$ shell closure and the entire fissioning region faster than in the baseline simulations. 
As can be seen from Figure~\ref{fig:beta_compare}, the neutron-induced fission flow increases due to the high flow of material above the third peak.
This effect is also noted and discussed in \citet{eichler2015}.
With less material accumulating in this region, particularly at $N=184$, the late-stage feeding of the actinides is significantly reduced. The lack of material at high $A$ is further illustrated in Figure~\ref{fig:avgA_marketin}, which shows the average mass number of the \rp\ material over time. 
Compared to the baseline case (Figure~\ref{fig:avgA_frdm2012}), the average $A$ is lower for the $Y_e^A$ and $Y_e^N$ simulations, most markedly at the robust fission recycling case, $Y_e^{N}$. 
Although fission cycles can still be identified in the patterns of Figures~\ref{fig:ye_marketin} and \ref{fig:avgA_marketin}, the high fission flows result in diminished variations relative to the baseline case.
Similarly, the abundance patterns at different values of $Y_e$ (Figure~\ref{fig:ab_marketin}) show little variation once fission begins to take place.

The faster \citet{marketin2016} $\beta$-decay rates near $A=130$ and above $A=190$ result in simulations that produce more europium and fewer actinides, and thus the predicted Th/Eu and U/Eu initial production ratios are significantly lower than for the baseline case. 
The impact of these ratios on the ages of \rp\ material in metal-poor stars is discussed in Section \ref{sec:prs}.


\subsection{Dependence on Mass Model: Duflo-Zuker}
\label{sec:dz33}

The \rp\ takes place far from stability where we must rely on nuclear mass models to estimate nuclear data. Nuclear masses set the reaction rate and decay $Q$-values required for calculations of all unknown reaction and decay properties. 
To study the effect of the mass model choice on predicted actinide production, we implement the Duflo-Zuker mass model with 33 terms \citep[DZ;][]{duflo1995}. We re-calculate all neutron-capture rates, photodissociation rates, $\beta$-decay half-lives, and $\beta$-delayed neutron emission probabilities using DZ masses as described in Section~\ref{sec:calculations} and \citet{mumpower2015}.
We continue to use \citet{moller2015} fission barrier heights, \citet{Moller+18} $\beta$-decay strength functions, and all experimental data.
Fission distributions and initial seed nuclei distributions also remain the same as in the baseline case.

\begin{figure}[t]
 \centerline{\includegraphics[width=0.95\columnwidth]{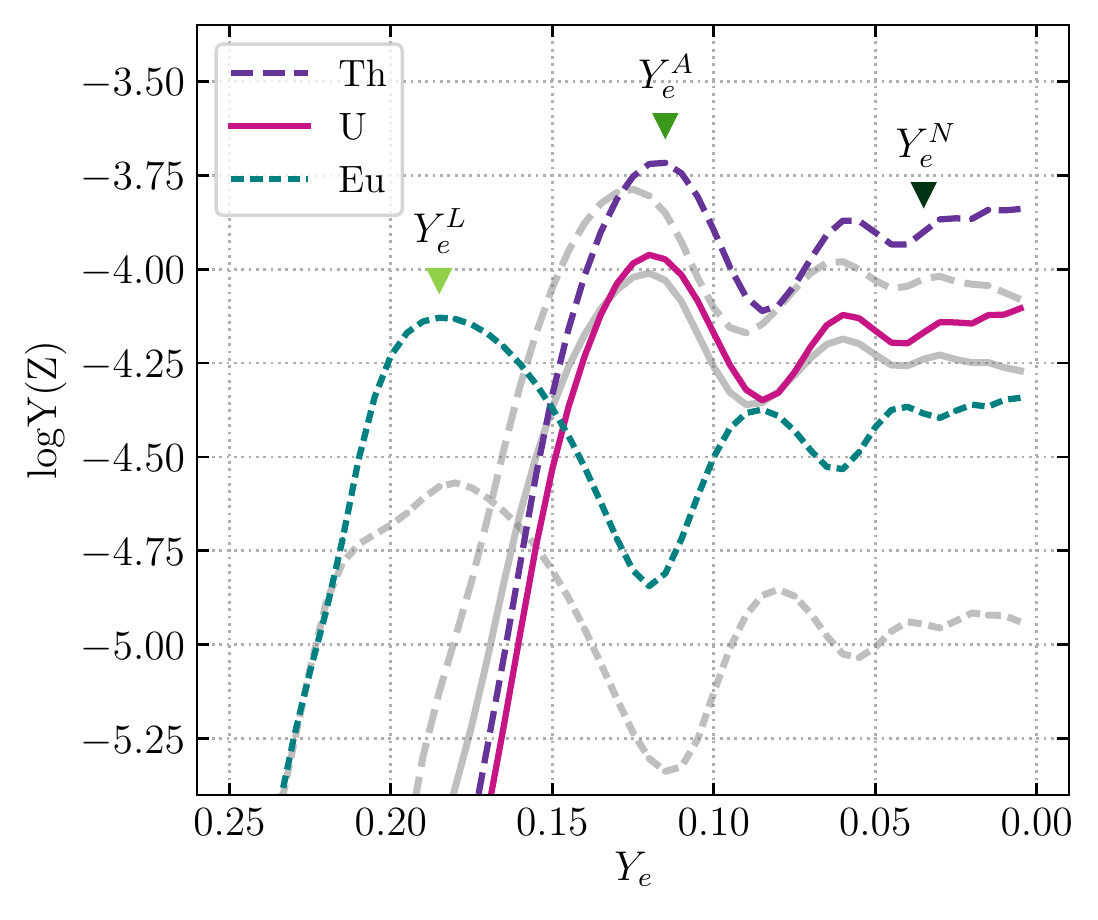}}
 \caption{\label{fig:ye_dz33} Europium, thorium, and uranium production as a function of initial $Y_e$ using 50/50 fission fragment distributions and the DZ mass model. The baseline case abundances of Figure~\ref{fig:ye_frdm2012} are plotted in gray for comparison.}
\end{figure}

Figure~\ref{fig:ye_dz33} shows the effect of varying initial $Y_e$ on Th, U, and Eu production in simulations using the DZ mass model.
Although the patterns of actinide and lanthanide production as a function of $Y_e$ remain similar to the baseline (FRDM2012) case, there are several features worthy of discussion.
Of particular note is the strong amount of Eu that is produced compared to the baseline calculations.

\begin{figure}[t]
 \centerline{\includegraphics[width=\columnwidth]{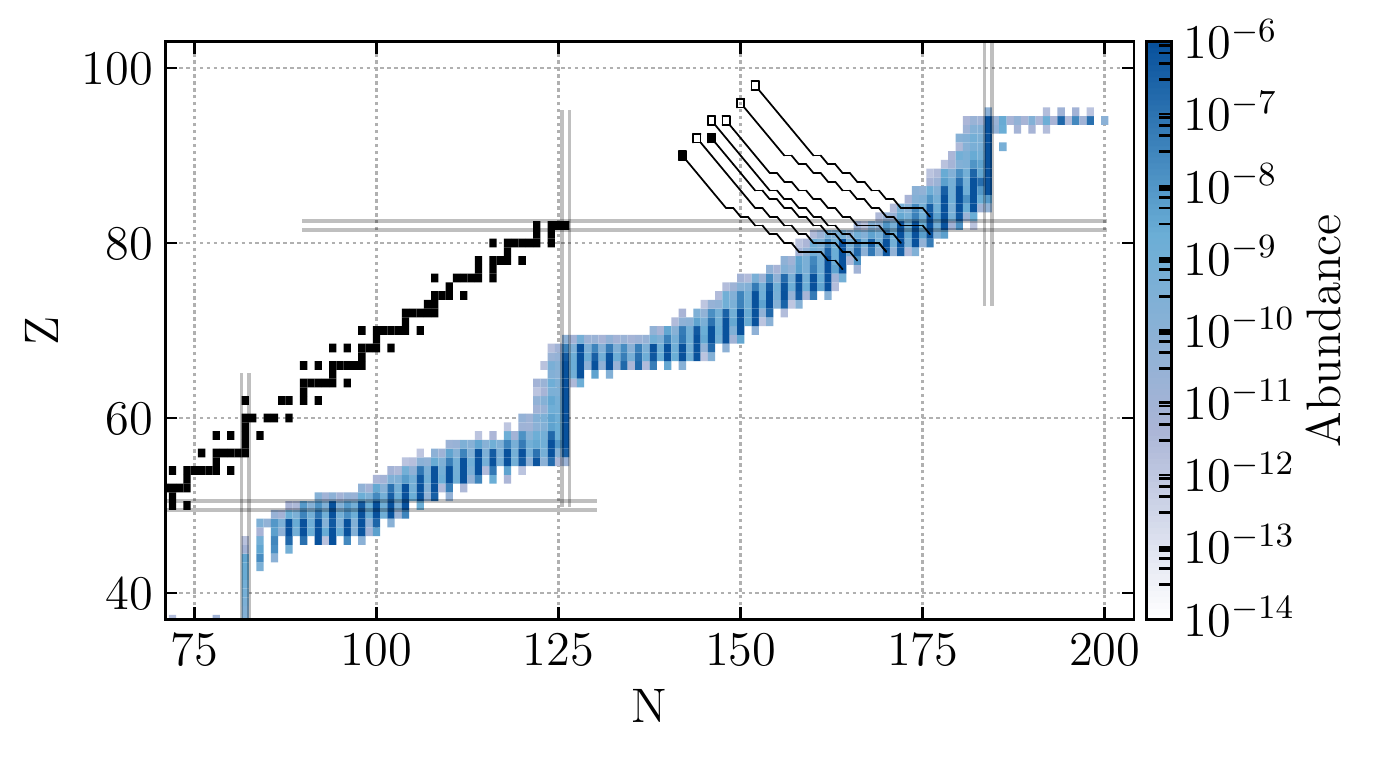}} 
 \centerline{\includegraphics[width=\columnwidth]{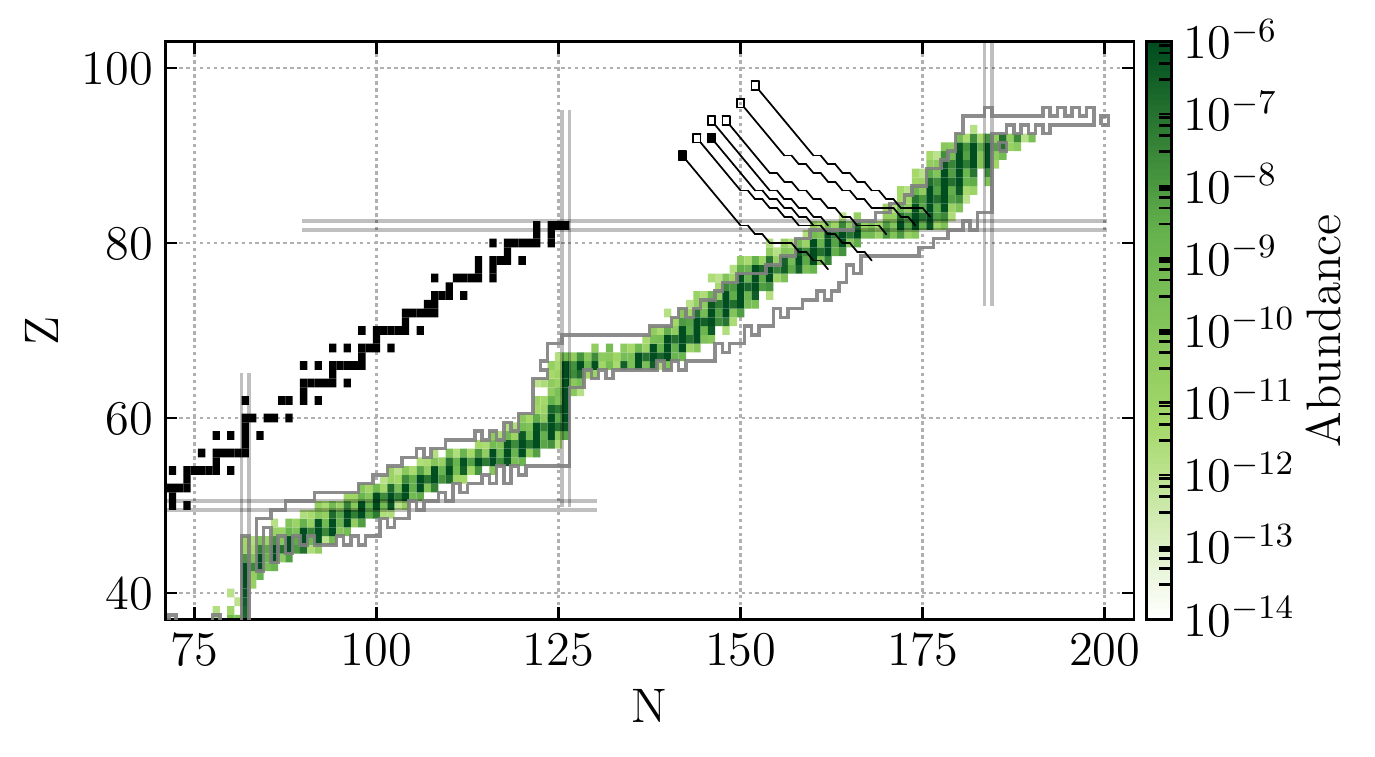}} 
 \caption{\label{fig:rpath_dz33} Abundances at feezeout using the baseline model (top) and the DZ mass model (bottom) at $Y_e=0.035$. The baseline FRDM2012 abundances are outlined in gray in the bottom panel for direct comparison.}
\end{figure}

\begin{figure}[t]
 \centerline{\includegraphics[width=\columnwidth]{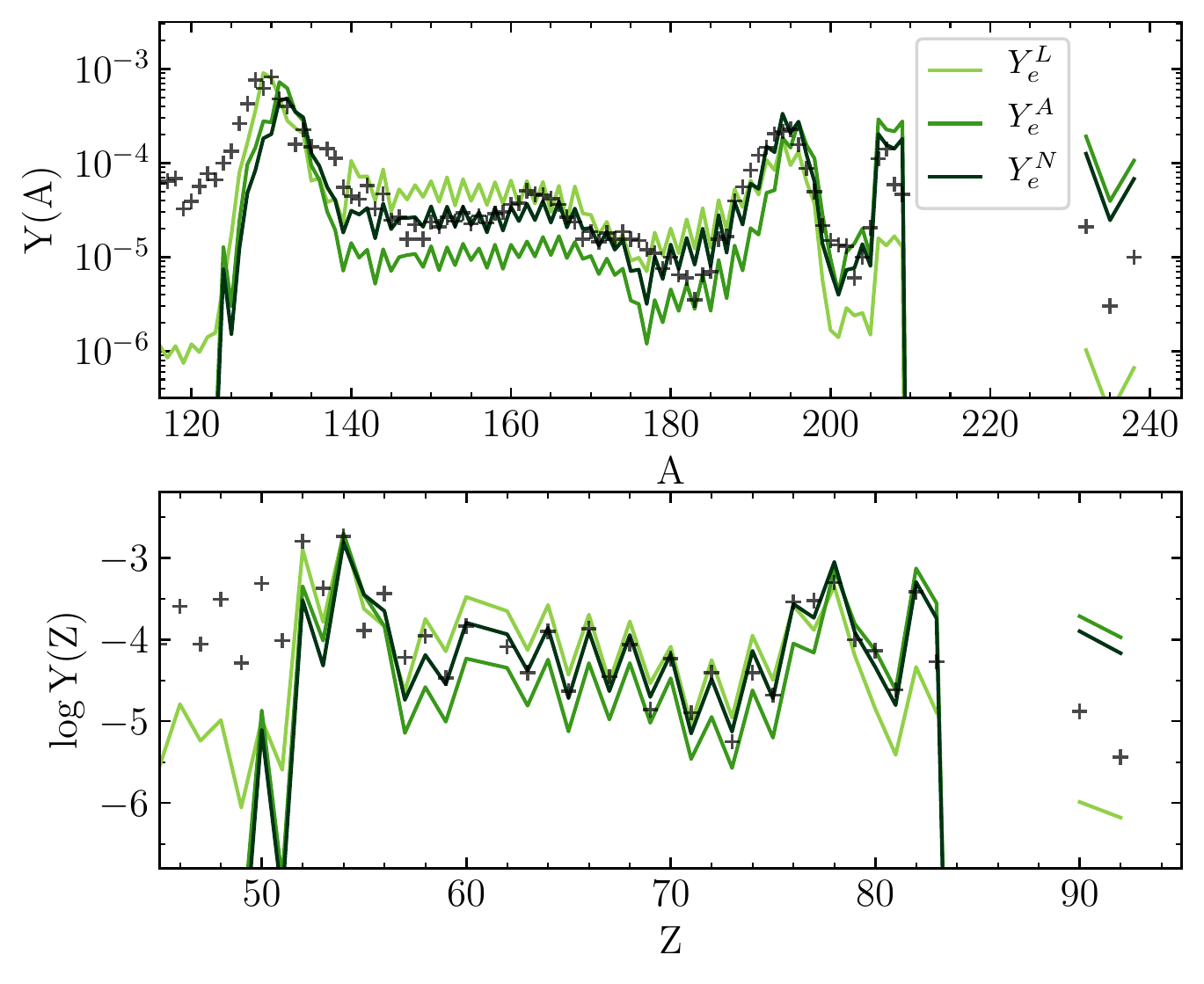}}
 \caption{\label{fig:ab_dz33} Final isotopic (top) and elemental (bottom) abundance patterns for the DZ case corresponding to the three initial $Y_e$ values where lanthanides are maximized ($Y_e^L$), actinides are maximized ($Y_e^A$), and the nominal value ($Y_e^N$). Scaled-Solar \rp\ abundances (``+") are from \citet{sneden2008}.}
\end{figure}

The one-neutron separation energies of DZ are generally lower than FRDM2012; the strength of the second \rp\ peak in FRDM2012 combined with the 50/50 fission distribution set leads to a characteristic deficiency just beyond the second peak, near $A=145$ \citep{kratz2014}.
DZ predicts somewhat weaker closed shells than FRDM2012, and as a result \rp\ simulations with DZ masses do not produce this deficiency. Instead, material flows smoothly and quickly to heavier masses, spending less time at the neutron shell closures. Figure~\ref{fig:rpath_dz33} shows the \rp\ path just before freezeout for both the baseline and DZ cases, using $Y_e=Y_e^N=0.035$.
Note that while overall the DZ \rp\ path sits slightly closer to stability, the ``kinks" in the path at the neutron shell closures are less pronounced than with the FRDM2012 masses. The \rp\ waiting points at the top of each closed shell are therefore a bit farther from stability in the DZ simulation and have shorter half-lives, leading to a reduced pile-up of material at the closed shells. This produces a final abundance pattern with a lower $A\sim 130$ peak and higher rare-earth region compared to simulations with FRDM2012, as shown in Figure~\ref{fig:ab_dz33}. 

The second notable feature is that actinide production requires lower $Y_e$ with DZ masses than with FRDM2012, as seen by the shifted rise in Th and U abundances in Figure~\ref{fig:ye_dz33} to lower $Y_e$.
This shift indicates that more neutrons are required to initiate actinide production. This is again because of the weaker shell structure of DZ compared to FRDM2012.
With DZ, less material is held up at the $N=82$ closed shell, so more $N>82$ material is able to capture neutrons and reach heavier masses.
With more material involved in capturing neutrons, a smaller fraction of those neutrons is available to populate the actinides.

For simulations sufficiently neutron-rich for actinide production and fission cycling, the actinide abundances shown in Figure~\ref{fig:ye_dz33} follow roughly the same trend with $Y_e$ as the baseline case.
One notable difference is that, at the extremely low $Y_e$ end, Th and U production steadily increase with decreasing $Y_e$, whereas they decrease in the baseline case.
At low $Y_e$, there is more neutron-induced fission when using FRDM2012 masses, which causes the divergence of the actinide abundances between the two models.
The higher separation energies of FRDM2012 compared to the incident neutron energy result in an increased likelihood to fission rather than neutron capture.

Simulations with the DZ mass model differ from the baseline most markedly in the production of europium.
Consequently, the Th/Eu and U/Eu production ratios are lower than the baseline at every $Y_e$ considered in this work.
The impact of these chronometers to the ages of \rp\ material is discussed in Section~\ref{sec:prs}.


\section{Production Ratios}
\label{sec:prs}


\subsection{Comparison to Observations}

We now apply the production ratios from Section~\ref{sec:production} to calculate the age of an \rp-enhanced, metal-poor star exhibiting an actinide boost, in order to examine whether the low-entropy dynamical ejecta from NSM could be the source of this signature.
We choose to consider the recently discovered \rii\ star J0954$+$5246 \citep{holmbeck2018}, which exhibits the strongest actinide boost of any metal-poor star studied to date. 

\begin{figure}[t]
 \centerline{\includegraphics[height=2.5in]{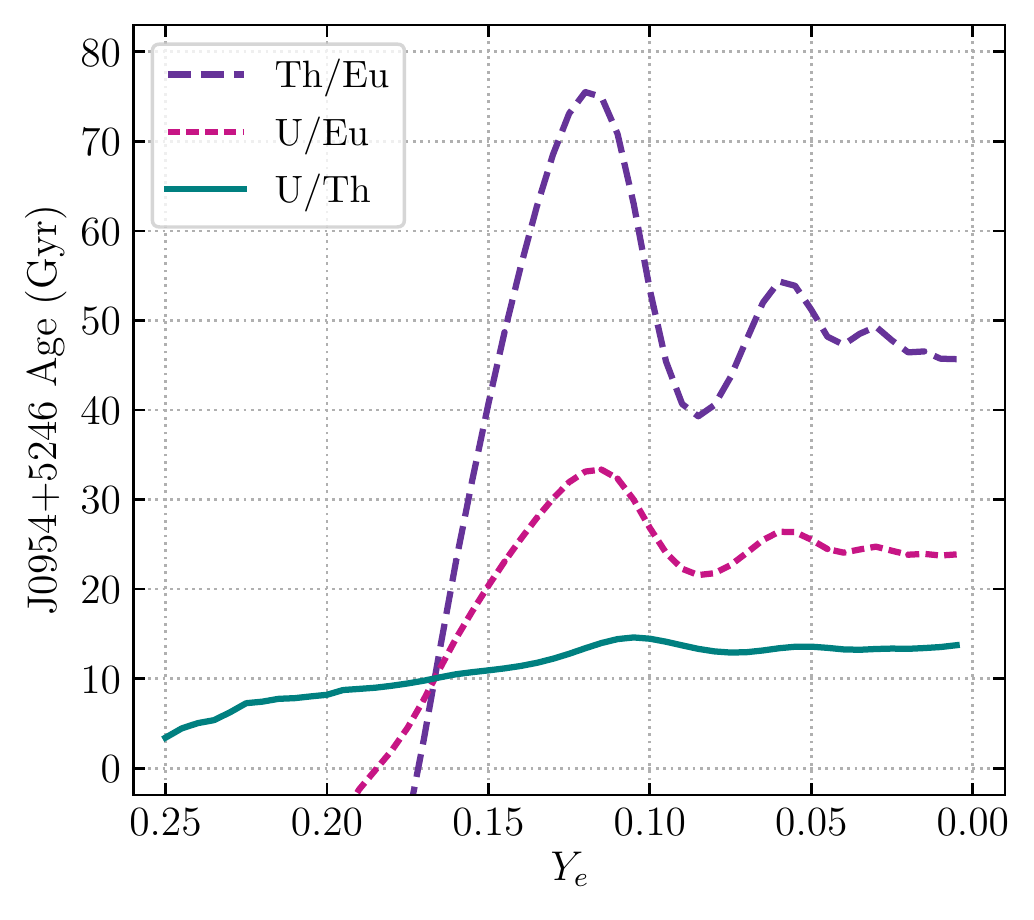}} 
 \caption{\label{fig:age_frdm2012} Age of J0954$+$5246 as a function of $Y_e$ using the baseline FRDM2012 mass model.}
\end{figure}

Figure~\ref{fig:age_frdm2012} shows the derived ages of J0954$+$5246 using the baseline Th, U, and Eu final abundances from Figure~\ref{fig:ye_frdm2012}.
These ages are calculated by applying the Th/Eu, U/Eu, and U/Th abundance ratios to Equations~\ref{eqn:theu}--\ref{eqn:uth}.
At sufficiently low $Y_e$, the U/Th production ratio stabilizes to a roughly constant value, and the age is estimated to be between 12 and 13~Gyr. However the Th/Eu and U/Eu ages are inconsistent with the U/Th age.
This inconsistency at every $Y_e<0.17$ suggests that if \rp\ conditions are very neutron-rich, the actinides are always over-produced, predicting a Th/Eu age that is far too high compared to observations of actinide-boost stars.

\begin{figure}[t]
  \centerline{\includegraphics[height=2.5in]{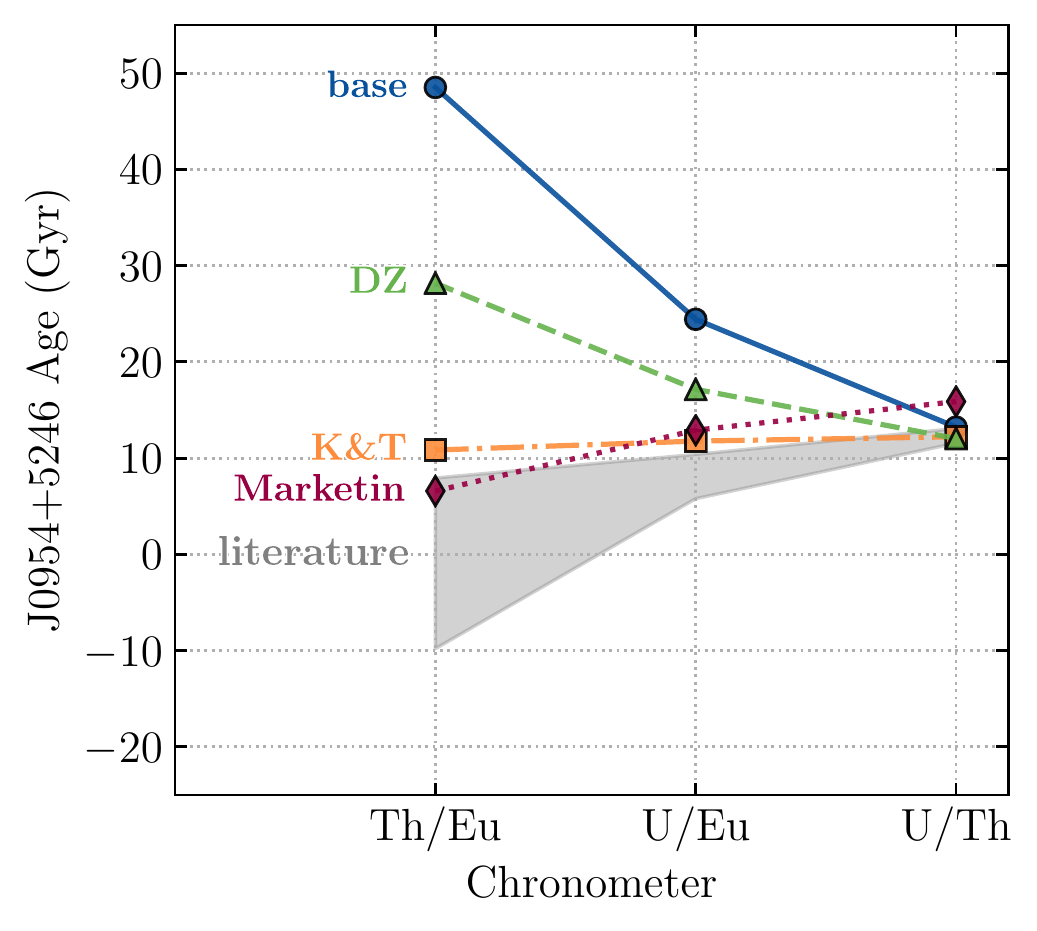}}
 \caption{\label{fig:pr_ages} Predicted age of the actinide-boost star J0954$+$5246 based on the three chronometer pairs Th/Eu, U/Eu, and U/Th for each case discussed. Production ratios are taken at $Y_e^N$.}
\end{figure}

As illustrated in Section~\ref{sec:production}, the predicted production ratios exhibit strong sensitivity to unknown nuclear physics properties of neutron-rich nuclei, and in particular Eu appears to be underproduced relative to Solar in the low-$Y_{e}$ baseline simulations. We therefore repeat the age estimates for J0954$+$5246 using production ratios calculated using each set of nuclear physics considered in Section~\ref{sec:production}. Figure~\ref{fig:pr_ages} shows the ages resulting from the $Y_e^N=0.035$ production ratios using all four cases studied in Section~\ref{sec:production}. For the model to successfully describe the observed ratios of \rp\ material present in the star, all three ages must agree (i.e., lie on a flat line). However, neither literature nor any case presented in Section~\ref{sec:production} succeeds in describing the actinide-boost star, J0954$+$5246. The cases range from underproducing the actinides (e.g., literature and Marketin) to overproducing them (e.g., base and DZ).

A variety of literature (SNe) production ratios \citep[e.g.,][]{schatz2002,wanajo2002,farouqi2010} all underproduce the actinides in J0954+5246, as shown in Figure~\ref{fig:pr_ages} by the extremely low (often negative) derived Th/Eu and U/Eu ages.
Negative ages result from simulation abundances that are less than observed abundances, implying negative (unphysical) actinide decay.
In contrast, the baseline and DZ models both overproduce the actinides, resulting in calculated Th/Eu and U/Eu ages that are much larger than the age of the Universe \citep[13.8 Gyr;][]{planck}. 
The simulation using the \citet{marketin2016} $\beta$-decay rates results in Th/Eu and U/Eu ages that follow a similar trend to those of the literature values. However, the U/Th ratio in this case is quite high and also suggests an age longer than the age of the Universe for both actinide-boost and non-actinide-boost stars.

The most consistent ages are obtained from the simulation with the \kt\ fission distributions. This result is misleading since both the actinides and europium are overproduced relative to the rest of the pattern, generating a smaller Th/Eu ratio. The actinides are still overproduced compared to lanthanides higher in $A$, which are less sensitive to fission fragment distribution than europium (e.g., Dy, $Z=66$, for which the Th/Dy age is $\sim$30 Gyr).
All but the Marketin case produce roughly similar U/Th ratios, which agree with published production ratios. No model, however, sufficiently describes the actinide-boost star J0954$+$5246.

It may be possible to produce both a match to the Solar elemental abundance pattern and consistent stellar ages with one fission-cycling trajectory through variations of the nuclear physics inputs beyond those considered here. As described in the previous section, actinide production relative to the rest of the main \rp\ is shaped by the nuclear masses, fission properties, and $\beta$-decay rates, while the europium abundance is particularly sensitive to the fission fragment distribution. Thus, consistent ages could be obtained by, for example, some combination of a different choice of fission fragment distribution in addition to changes to the $\beta$-decay or fission rates above $N=126$. More systematic studies of nuclear properties of heavy nuclei, particularly above $N=126$, by both theorists and experimentalists are clearly needed. Given the nuclear inputs as chosen, we turn our attention to whether an astrophysical solution can be found to yield consistency between the implied ages.


\subsection{The Actinide-Dilution Method}

Across a variety of nuclear physics inputs, actinides are generally predicted to be overproduced in the low-entropy dynamical ejecta with fission recycling, and none of the models succeed in providing both realistic production ratios and abundance pattern-matching for actinide-boost stars.
The low-entropy dynamical ejecta, however, are only one component in NSMs capable of harboring \rp\ nucleosynthesis.
Nucleosynthesis may also occur in the wind of a NSM (e.g., from the accretion disk), which is typically characterized by higher $Y_e$ and higher entropy than the astrophysical trajectory considered here.
If the disk wind produces a sufficiently robust \rp\ pattern (i.e., up to Eu) without synthesizing the actinides, then Th and U abundances can be diluted by the material from the NSM disk wind, lowering the total Th/Eu and U/Eu abundance ratios from the NSM event.

\begin{figure}[t]
 \centerline{\includegraphics[width=\columnwidth]{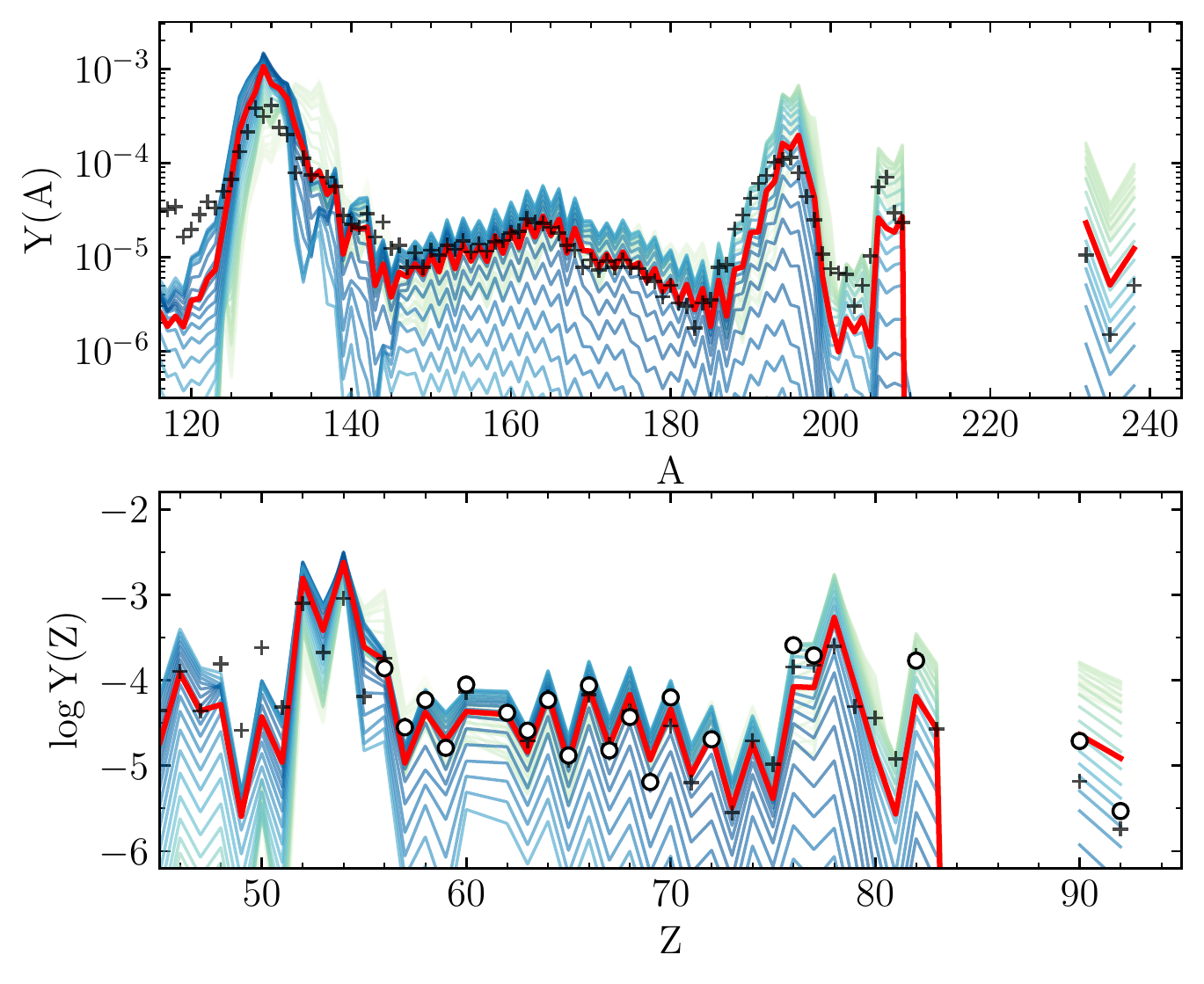}}
 \caption{\label{fig:ab_frdm2012_total} The individual (green/blue) and combined (red) abundance patterns that constitute the AD method applied to our baseline FRDM2012 case. Each individual pattern corresponding to different initial $Y_e$ is colored by their mass contribution with light green being the lowest and blue the highest. Scaled-Solar \rp\ abundances (``+") from \citet{sneden2008} and J0954+5246 (circles) are shown for comparison.}
\end{figure}

\begin{figure}[t]
 \centerline{\includegraphics[height=2.5in]{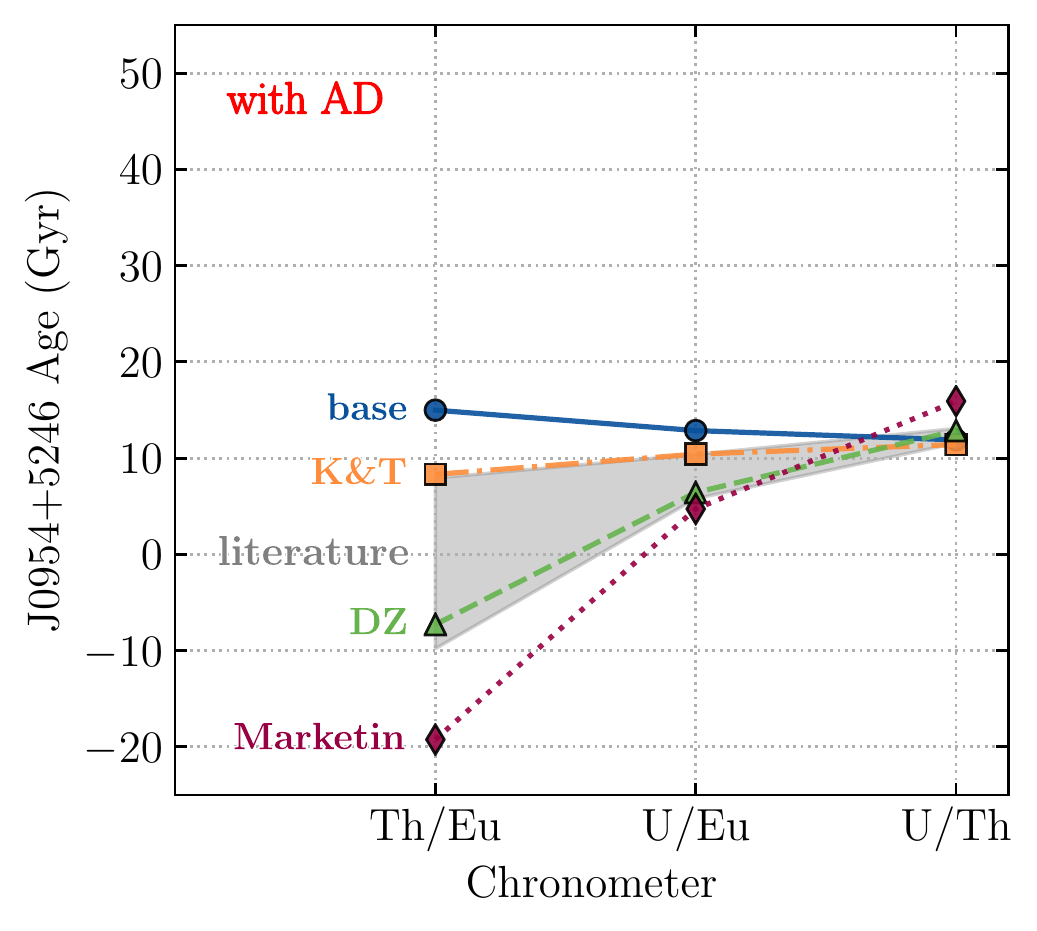}}
 \caption{\label{fig:pr_mix} Predicted age of actinide-boost star J0954$+$5246 based on the three chronometer pairs Th/Eu, U/Eu, and U/Th after applying the AD method to each case.}
\end{figure}

To estimate how combinations of NSM disk and dynamical nucleosynthesis can provide an explanation for the actinide boost, we present a simple actinide-dilution (``AD") method.
In this model, we start with mass distributions as a function of $Y_e$ from the literature: the low-entropy H000 model of \citet{lippuner2017} for the disk wind and the SFHO model of \citet{bovard2017} for the dynamical ejecta.
We fit Gaussian functions to these two distributions and obtain fits with centroids at $Y_e=0.16$ and 0.22 and one-sigma spreads of 0.042 and 0.037 for the dynamical ejecta and disk wind, respectively.
Next, we normalize the Gaussian fits to the ratio of the mass distributions of the dynamical ejecta ($m_{\rm dyn}$) and disk wind ($m_{\rm w}$).
We adopt $m_{\rm w}/m_{\rm dyn}=3$, which both \citet{rosswog2017} and \citet{tanaka2017} estimate as the ejected mass distribution ratio for GW170817.
This method leaves us with a double-Gaussian approximation for the mass distribution as a function of $Y_e$.
Finally, we multiply the final abundances of each $Y_e$ case by the mass fraction modeled by our fitted double-Gaussian distribution of ejecta.
Combining the adjusted abundances by this weighting scheme results in the abundance pattern of Figure \ref{fig:ab_frdm2012_total} using the baseline abundances.
Figure~\ref{fig:pr_mix} shows the predicted age of J0952+5246 using this weighting scheme applied to all cases considered in Section~\ref{sec:production}.
It is important to note that we have used neither the abundances of J0954+5246 nor the abundances calculated in Section~\ref{sec:production} to inform the choice of mass distribution.
Rather, the same fitted mass distribution is applied to all four cases.

The specific mass distribution assumed by the AD method are such that material which has undergone little to no fission cycling constitutes a majority of the ejecta mass. Therefore, the baseline case---which produced the most unrealistic X/Eu stellar ages when applied to one fission-cycling trajectory as in Figure~\ref{fig:pr_ages}---is more successful with AD. The overproduction of actinides and underproduction of europium characteristic of the baseline fission-cycling abundances are moderated by contributions at higher $Y_e$ that fill in the europium and dilute the actinides, resulting in abundance patterns that are a good match to Solar and J0954+5246, as shown in Figure~\ref{fig:ab_frdm2012_total}.

The DZ and Marketin cases are less successful with AD. Simulations using Marketin $\beta$-decay rates do not tend to overproduce actinides, even in fission-cycling conditions, and dilution results in extreme negative ages. The unrealistic U/Th age for the Marketin case in Figure~\ref{fig:pr_ages} is not influenced by actinide dilution, and this case is still ruled out under AD. On the other hand, the DZ case shows similar, if not as extreme, X/Eu age overestimates as the baseline case in Figure~\ref{fig:pr_ages}. The X/Eu ages become negative, however, under the specific choice of mass distribution and ratio of dynamical-to-wind mass ejecta mass used with the AD method in Figure~\ref{fig:pr_mix}. A slight modification to the mass choice used by the AD method could produce a consistent age for the DZ case.

The \kt\ case appears to be the most successful when applying AD. The reduction of fission-cycling material included in the chosen mass distributions mitigates the overproduction of both europium and actinides, leading to similar actinide-to-europium production ratios before and after AD. The different fission fragment distributions used by the baseline and \kt\ cases could perhaps represent extreme variations of fission yield asymmetry. Seeing as perfect consistency between the Th/Eu, U/Eu, and U/Th ages lies between the baseline and \kt\ cases with the AD method, modern treatments of fission fragment distributions \citep[e.g.,][]{goriely2013,eichler2015} could further improve our ability to reproduce observational data.

By simulating a wind component that adds a significant high-$Y_e$ contribution as suggested by the literature, the estimated actinide production in the low-entropy dynamical ejecta is sufficiently diluted to better account for the observations of J0954+5246.
 We postulate that the \rp\ event that produced the material observable in J0952+5246 could have been produced by a NSM event occurring 12.9 billion years ago.

Although $Y_e$ is not the only astrophysical parameter describing NSM ejecta, we are able to reproduce the observations of an actinide-boost star by adjusting relative contributions based only on the $Y_e$.
The actinide-boost phenomenon is observed over a range of enhancement levels in different \rp-enhanced stars; if the low-$Y_e$ dynamical ejecta of NSMs produce this actinide boost, different mixing ratios between the wind and dynamical ejecta components could account for the observed variation in Th/Eu levels in metal-poor stars.
Since actinides are grossly overproduced once fission cycling begins, the actinide-boost variation cannot be explained by the presence of fission cycling alone. Instead, we propose that the observed actinide variation results from the mixing ratio between {\rp es} with slightly different astrophysical conditions, namely $Y_e$.
Since J0954+5246 is presently the highest actinide-boost \rii\ star, the ratio we assume of $m_{\rm w}/m_{\rm dyn}\approx 3$ is simply a lower limit; more disk mass would be required to account for \rp-enhanced stars with lower Th/Eu ratios, as these observed ratios would require more dilution of the actinides.
Metal-poor stars with observed ratios of Th/Eu lower than that of J0954+5246 would require a higher mass ratio of disk wind to dynamical ejecta according to our actinide-dilution method.


\section{Conclusions}

The NSM dynamical ejecta are attractive environments for the production of lanthanides and actinides via \rp\ nucleosynthesis.
We consider wide variations in the initial neutron richness of the ejected material and distinct choices for the nuclear physics: fission fragment distributions, $\beta$-decay rates, and nuclear mass models. We find that the ratio of europium to uranium or thorium is quite sensitive to the electron fraction of the outflow, unless the $Y_{e}$ is sufficiently low to result in fission recycling, where roughly constant production ratios are achieved. These constant ratios almost uniformly predict an overproduction of actinides relative to lanthanides.

While the predicted overproduction of actinides points to a possible avenue to explain the phenomenon of actinide-boost stars, we find our NSM dynamical ejecta production ratios do not result in realistic age estimates for even the most actinide-boosted star discovered so far, J0952+5246. This may suggest that a significant disk wind component which synthesizes lanthanides but fewer actinides is required. We construct a simple model, combining low-$Y_{e}$ dynamical ejecta with higher-$Y_{e}$ disk material according to modeled $Y_{e}$ distributions from the literature. We find that a ratio of disk to dynamical ejecta mass of $\sim$3 produces realistic Th/Eu and U/Eu ages that are consistent with the U/Th age of an actinide-boost star. 

The range in actinide ratios observed thus far in \rp-enhanced metal-poor stars can possibly be explained as coming from neutron star mergers with varying ratios of disk to dynamical ejecta. In future work we will explore this possibility by extending our actinide-dilution (AD) model to include the full variety of disk and dynamical outflow trajectories from modern NSM simulations. 
As part of the $R$-Process Alliance (RPA) effort, much larger numbers of \rp-enhanced stars are currently being discovered in the Galactic halo \citep[e.g.,][Ezzedine, \emph{in prep}]{hansen2018,sakari2018}, including additional stars with both U and Th measurements, enabling a refined estimate of the fraction of actinide-boost stars and better constraints on the observed variation in the range of derived Th/Eu, U/Eu, and U/Th ratios among \rp-enhanced stars. 
The precision to which we can reliably quantify the disk/dynamical mass ratio necessary to explain observational data will still be limited by uncertainties in the nuclear physics, neutrino physics, and stellar astrophysics. In future work, we will aim to quantify the impact of these uncertainties as we look forward to experimental, observational, and theoretical progress in these areas.

\acknowledgments{
We thank the anonymous referee for their constructive feedback that improved the quality of our manuscript.
E.M.H., R.S., and T.C.B. acknowledge partial support for this work awarded by the US National Science Foundation grant PHY 14-30152; Physics Frontier Center/JINA Center for the Evolution of the Elements (JINA-CEE).
R.S., M.R.M., N.V., and T.K. were supported in part by the U.S. Department of Energy (DOE) under Contract No. DE-AC52-07NA27344 for the Fission In R-process Elements (FIRE) collaboration.
This work was partly supported by the U.S. DOE under Award Numbers DE-SC0013039 (R.S. and T.M.S.) and DE-SC0018232 (T.M.S.). 
A portion of this work was carried out under the auspices of the National Nuclear Security Administration of the U.S. DOE at Los Alamos National Laboratory (LANL) under Contract No. DE-AC52-06NA25396 (M.R.M. and T.K.)}

\bibliography{bibliography.bib}


\end{document}